\documentclass[twocolumn,tighten]{aastex701}

\usepackage{booktabs}
\begin{document}

\title{Double Dots: Compact Pairs Mark Little Red Dots and High-Redshift Broad-line AGNs}

\correspondingauthor{Amy J. Barger}
\email{barger@astro.wisc.edu}

\author[0000-0002-3306-1606]{A.~J.~Barger}
\affiliation{Department of Astronomy, University of Wisconsin-Madison,
475 N. Charter Street, Madison, WI 53706, USA}
\affiliation{Department of Physics and Astronomy, University of Hawaii,
2505 Correa Road, Honolulu, HI 96822, USA}
\affiliation{Institute for Astronomy, University of Hawaii, 2680 Woodlawn Drive,
Honolulu, HI 96822, USA}
\email{barger@astro.wisc.edu}

\author[0000-0002-6319-1575]{L.~L.~Cowie}
\affiliation{Institute for Astronomy, University of Hawaii,
2680 Woodlawn Drive, Honolulu, HI 96822, USA}
\email{cowie@ifa.hawaii.edu}

\newcommand{\alma}{870\,$\mu$m}
\newcommand{\md}{\,M$_\odot$}
\newcommand{\sfr}{M$_\odot$\,yr$^{-1}$}
\newcommand{\fluxunits}{\,erg\,cm$^{-2}$\,s$^{-1}$}
\newcommand{\powerunits}{\,erg\,s$^{-1}$}
\newcommand{\esHz}{\,erg\,s$^{-1}$\,Hz$^{-1}$}

\begin{abstract}
We utilize JWST imaging of the massive lensing cluster field A2744 to find close pairs
of compact sources with separations less than $0\farcs25$. 
A large fraction of these
 ``Double Dots" correspond to Little Red Dots (LRDs) or high-redshift broad-line active
 galactic nuclei (BLAGNs).
Our analysis of 31 identified pairs reveals a median separation of $0\farcs15$. 
Statistical comparison against a 
uniform  background shows that these are mostly physical pairs.
We find that at least 16 of the 24 previously published LRDs in this field 
($\sim67$\%) are such pairs, as are both of the high-redshift BLAGNs.
We demonstrate that the presence of a companion can significantly contaminate 
the measured spectral energy distribution, potentially 
masking the characteristic ``v-shape" used for LRD classification. 
Furthermore, our 2D spectroscopic analysis of several pairs reveals that BLAGN activity 
is not confined to the redder member of the pair but can originate in either one. 
Since most LRDs contain broad emission lines, our findings suggest that close pairs are 
extremely effective markers of galaxies with broad lines at high redshift.
We speculate on possible mechanisms, concluding that we are likely
seeing merger-driven accretion.
\end{abstract}

\section{Introduction}
\label{sec:intro}
JWST observations have shown that broad-line sources are surprisingly common at high redshift ($z=4-7$). 
This includes both blue broad-line active galactic nuclei (BLAGNs; \citealt{harikane23}) and Little Red Dots  
(LRDs; \citealt{matthee24,greene24,labbe25,kocevski25}),
most of whose spectra show broad-line H$\alpha$ (e.g., \citealt{taylor25}).

Currently, the favored interpretation of the LRDs is
that they are high-redshift supermassive black holes, as supported by the detection of the
broad Balmer emission lines.  
There are still many questions about this interpretation, however.
Their black hole masses would be only $10^{6}$ to $10^{8}$\,\md, which is orders of magnitude lower in mass 
than the luminous quasars accessible before 
JWST \citep{harikane23,maiolino24,matthee24}. 

In addition to their abundance, LRDs have very unusual properties: they are systematically overmassive 
relative to their host galaxies based on
local scaling relations, and they are strikingly weak in X-ray, mid-infrared, submillimeter,
and radio emission \citep{ananna24,barger26,mazzolari26}. The LRDs appear
to represent the higher luminosity end of the BLAGN population (\citealt{taylor25}), 
though whether they are physically distinct or simply the most obscured examples remains unclear.

Indeed, driven by these problems, alternatives to the BLAGN picture continue to be put forward.  
Most recently, \cite{chisolm26} have proposed that LRDs may instead be globular clusters in formation, 
with the rest-frame UV arising from a very young stellar population, 
and the rest-frame optical dominated by short-lived supermassive stars.

Accepting the BLAGN interpretation, we still need very unusual accretion properties to account for the observations
and high-mass seeds to explain their overmassive nature. 
The lack of significant X-ray emission suggests they may be in a super-Eddington accretion regime where  
``slim" accretion 
disks with high-density coronae can lead to softer X-ray spectra and suppressed coronal emission (\citealt{pacucci24}).
The characteristic ``v-shape" spectral energy distribution (SED) has been  interpreted as the signature of  dense gas 
enshrouding the  AGN \citep{greene24,matthee24}. The 
high masses may be explained by the Direct Collapse Black Hole (DCBH) model
\citep{loeb94,volonteri05,begelman06,lodato06,lodato07},
where the high-mass
seeds can form  when the Lyman-Werner cooling is suppressed by the radiation from a neighboring galaxy
(\citealt{pacucci26}).
 This would require the presence of a companion object irradiating the collapsing cloud,
a configuration that would predict a high incidence of close pairs \citep{baggen26}. 

Nearly all of the possibilities we have discussed above, including merger-driven accretion,
direct black hole formation, and globular cluster formation would likely imply that 
the objects should have complex substructure. Treating them as single objects 
may just reflect the  key limitation of spatial resolution.
At the angular scales probed by the F444W band of JWST, what appears to be a single compact source
may in fact be two or more physically associated objects blended together. If high redshift broad-line sources  and LRDs 
frequently contain close pairs,
the implications would be significant: their SEDs would be contaminated by companion light, 
their photometric classifications would be systematically biased, and the spectroscopic signatures 
used to identify broad-line  activity might be attributed to the wrong component. 

In this paper, we exploit the exceptional depth of JWST imaging and enhanced resolution in the lensing cluster 
field A2744 (T. Treu et al. 2022; R. Bezanson et al. 2024; J. R. Weaver et al. 2024)
 to search systematically for such close pairs among compact sources. We identify 31 compact pairs — which
 we term "Double Dots" — with a median (not delensed)  separation of just 0.15 arcseconds. 
We find that at least 16 of the 24 previously published LRDs in this 
field (J. E. Greene et al. 2024; I. Labbe et al. 2025; D. D. Kocevski et al. 2026) correspond
 to such pairs, and we demonstrate through SED decomposition and 2D spectroscopic analysis
that companion contamination has significant effects. Strikingly, and consistent with the 
broader JWST picture of high-redshift broad-line AGN (Y. Harikane et al. 2023), the AGN 
activity in these pairs is not confined to the redder component, suggesting that close 
pairs may serve as a fundamental marker of high-redshift AGN.

The paper is structured as follows. In Section~2, we describe our selection. 
In Section~3, we present our analysis of the SEDs and spectra in the pairs.
Finally, in Section~4, we summarize our results and consider possible interpretation.

We assume a standard cosmology of $H_0=70.5$~km~s$^{-1}$~Mpc$^{-1}$,
$\Omega_{\rm M}=0.27$, and $\Omega_\Lambda=0.73$ \citep{larson11}. In calculating
masses and SFRs, we assume a \cite{kroupa01} initial mass function.

\section{Data}
\label{sec:data}

\subsection {JWST imaging}
\label{subsec:JWST}
The A2744 field has JWST observations from several surveys, including the GLASS-A2744 \citep{treu22} 
and UNCOVER \citep{bezanson22} programs.  Here we use the DR2 photometric catalog
from \citet{weaver24}, which contains additional data products provided 
by the UNCOVER team, including lensing magnifications, photometric and 
spectroscopic redshifts, and rest-frame HST and JWST fluxes in many bands. 
The provided lensing magnifications were derived from the lensing model of \citet{furtak23b}, 
which is highly constrained using HST and JWST multiply-imaged sources. 

We started by identifying
moderately compact sources (FWHM $<$0\farcs2 in F444W) with 
S/N $>20$ in the F444W band, which were not identified as 
stars. The FWHMs are the observed values
without any deconvolution of the point spread function
or correction for lensing. After a first-pass 
visual inspection to confirm the sources were real,
we compiled a list of 153 sources. 
This list contains all 24 published LRDs in the field
(\citealt{greene24,labbe25,kocevski25}), excluding one brown dwarf and one 
source with S/N below 20  in the F444W band.

Next, we visually identified which of these sources are double in 
F150W, which we define here as having two or more sources in a $0\farcs25$
radius region (31 pairs; see Table~\ref{tab:full_data} where we give
the coordinates for all of the pairs).
We refer to this as {\em our compact pairs sample} 
and label the individual objects DD0 to DD61.
These sources are generally unresolved in F444W and appear
as a single compact source at this wavelength (e.g. Figure~\ref{fig:doublered}).
Eleven of the 24 published LRDs, as well as the two 
BLAGNs found in the field by \citet{harikane23}
are contained in this sample, as noted in the table.
As noted in the Introduction, since most of the LRDs contain broad lines,
it appears that the close pairs are extremely effective
markers of high-redshift broad line sources in the field.

\begin{figure}[t]
\includegraphics[width=3.25in,angle=0]{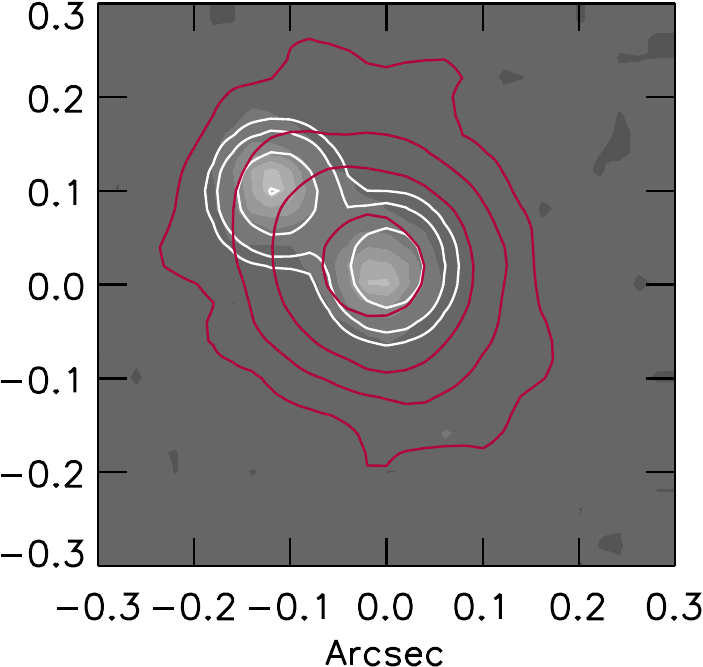}
\caption{Illustration of the fitting procedure. We show the
F150W image of the pair DD20 and DD21. Overlaid (white contours)
are the two 2D Gaussian fits to the F150W image. Each has a fitted
$x$ and $y$ width; in this case, the fits are near circular.
We use the geometric mean of the $x$ and $y$ measurements for each 
member of the pair as the FWHM of that component.
Here, both components have FWHM $=$ 0\farcs083
(see Table~\ref{tab:full_data}). Also overlaid is the F444W image (red contours),
illustrating how strongly concentrated the F444W light is on the lower 
right member of the pair.
\label{fig:fit_example}
}
\end{figure}

We measured  the source size of each member of
a pair using two 2D Gaussians fit to the F150 image
using the IDL MPFIT routine \citep{markwardt09}.
We illustrate the procedure in Figure~\ref{fig:fit_example}.
For each member of the pair, we fit  a 2D
Gaussian with separate measures of the x and y widths.
We take the FWHM to be the geometric mean of the FWHM 
measured in each of the two axes. This leaves us with
measurements of the sizes of the two members of the pair 
(see Table~\ref{tab:full_data}).
(The FWHM of the PSF at this wavelength is $0\farcs05$,
though it is slightly undersampled in the NIRCAM image). 

\begin{figure}[t]
\includegraphics[width=3.25in,angle=0]{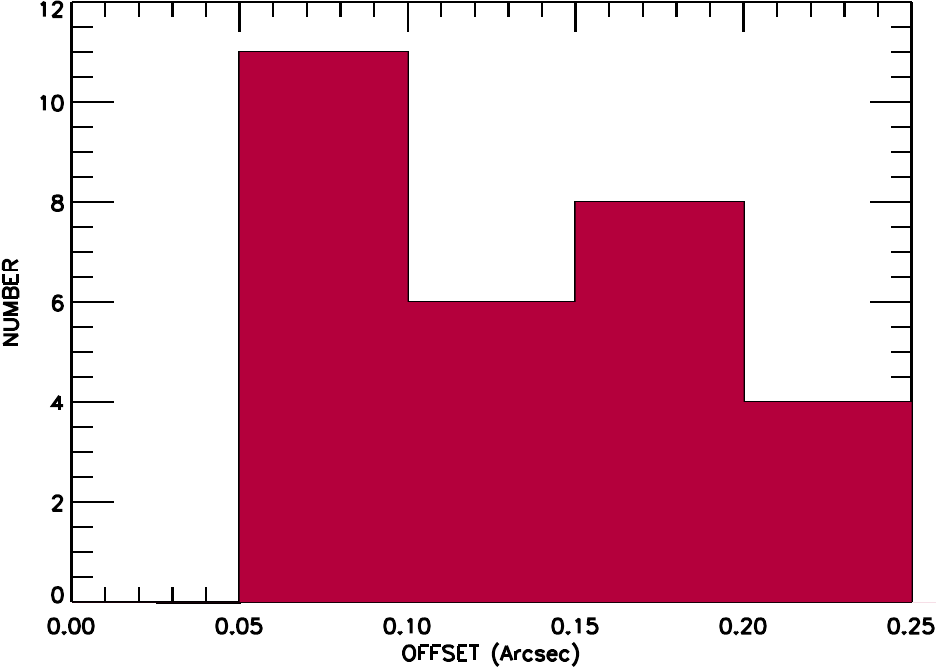}
\caption{The distribution of offsets between the pairs.
We cannot measure offsets smaller than 0\farcs05
given the spatial resolution in the F150 image. However,
at larger separations, the distribution is skewed to smaller values,
which is not consistent with expectations for a background
population.
\label{fig:offset-hist}
}
\end{figure}

The double fitting also provides an accurate measure of the pair
separation. The median separation is $0\farcs15$, though
one of the sources has an offset slightly larger
than $0\farcs25$. The distribution is skewed to smaller separations
(Figure~\ref{fig:offset-hist}),
suggesting that these are physical pairs rather than chance projections.
A Chi-Square  analysis comparing
with the values expected from a uniform background model 
gives a p-value of only  $5.2\times10^{-7}$ strongly rejecting
the hypothesis that the sources are drawn from the  background population.

\begin{figure}[t]
\includegraphics[width=3.4in,angle=0]{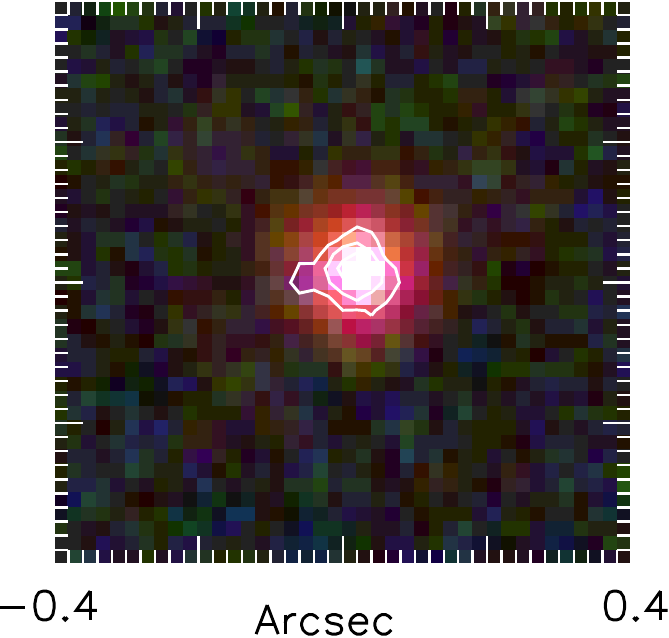}
\caption{RGB images (F444W, F200W, and F150W) for
an LRD (R.A. 3.5694625, Decl. -30.339306)
from \citet{labbe25}.
Overlaid (white contours) is the combined F115W and F150W image. 
The companion is 0\farcs05
separated from the peak and is just below the limit where it would
be included in our pairs sample.
\label{fig:close_pair}
}
\end{figure}

\begin{figure}[t]
\includegraphics[width=3.4in,angle=0]{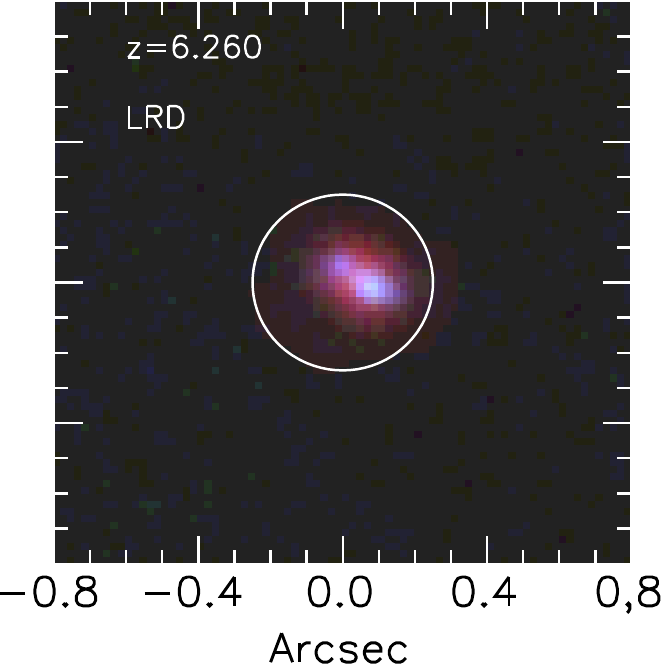}
\caption{
RGB images (F444W, F200W, and F150W) for the one pair in our sample
(DD54 and DD55) that consists of two red sources.
White circle shows a 0\farcs25 radius. 
\label{fig:doublered}
}
\end{figure}

The steep rise to smaller separations suggests that there may be LRDs with
closer separations than would be included in our pairs sample. We therefore
searched the full LRD sample in a combined F115W and F150W image
to see if cases with more closely separated pairs might be found. A further
5 objects have companions with closer separations. We show one example
in Figure~\ref{fig:close_pair}. In total, at least 16 of the 24 previously published
LRDs in this field are pairs.

\begin{figure*}[t]
\includegraphics[width=2.25in,angle=0]{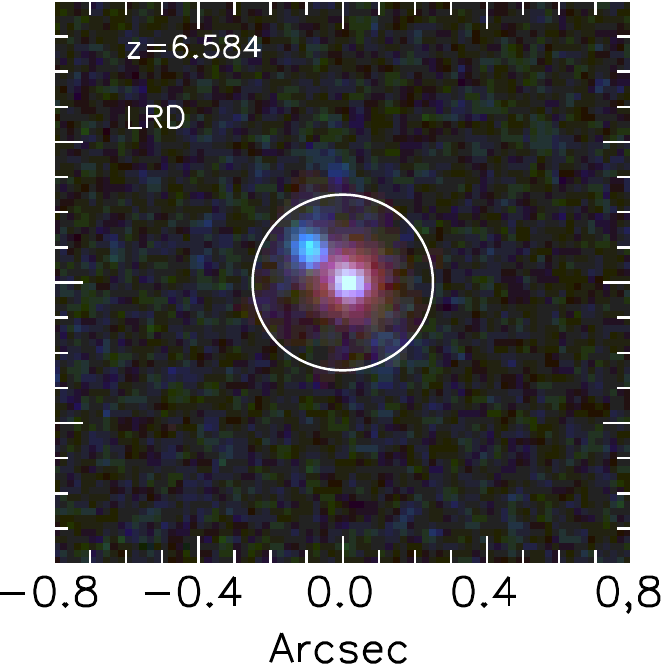}
\hspace{-2cm}\includegraphics[width=2.25in,angle=0]{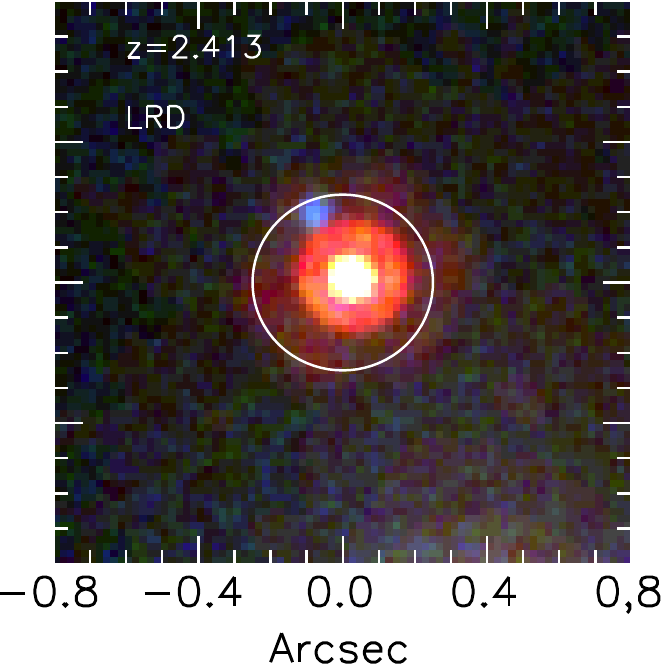}
\hspace{-2cm}\includegraphics[width=2.25in,angle=0]{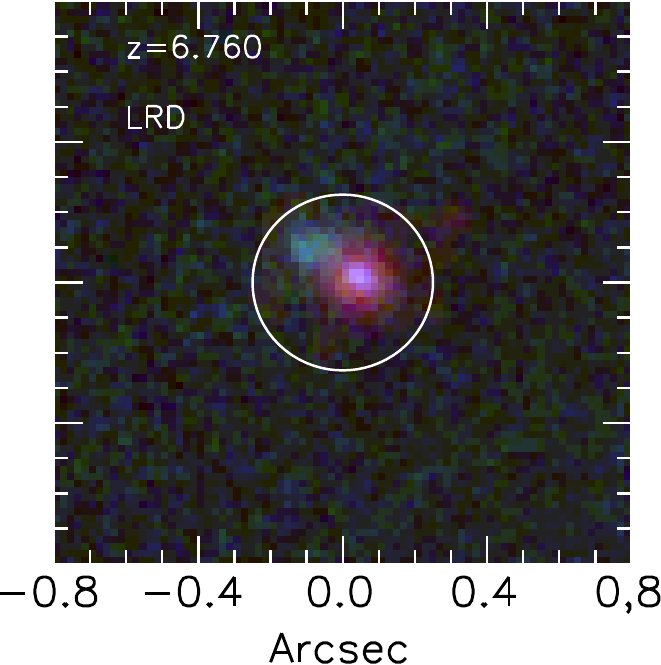}
\includegraphics[width=2.25in,angle=0]{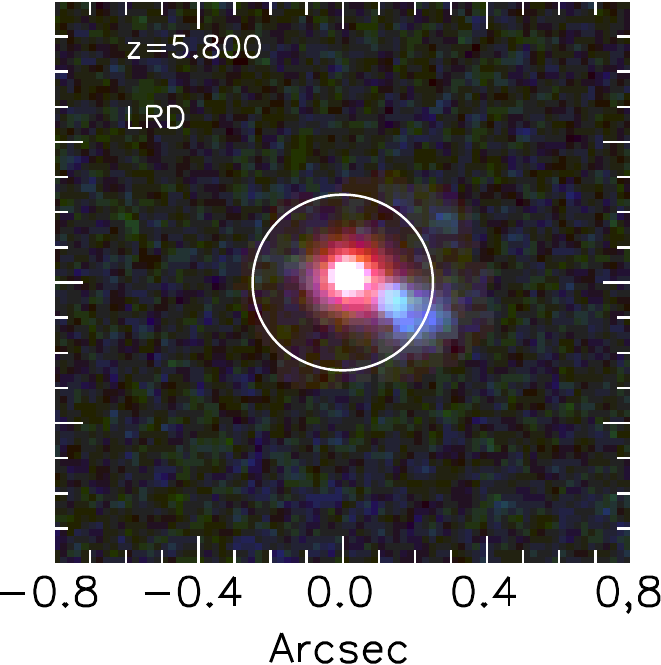}
\hspace{0.3cm}\includegraphics[width=2.25in,angle=0]{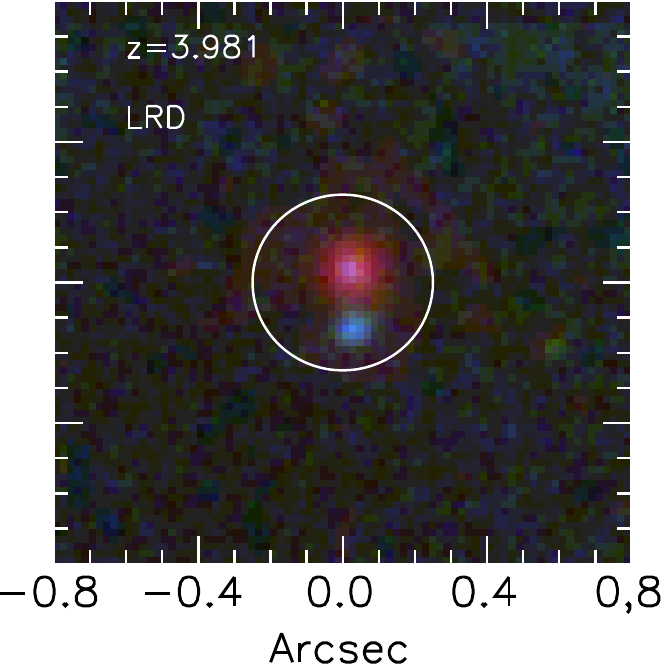}
\hspace{0.3cm}\includegraphics[width=2.25in,angle=0]{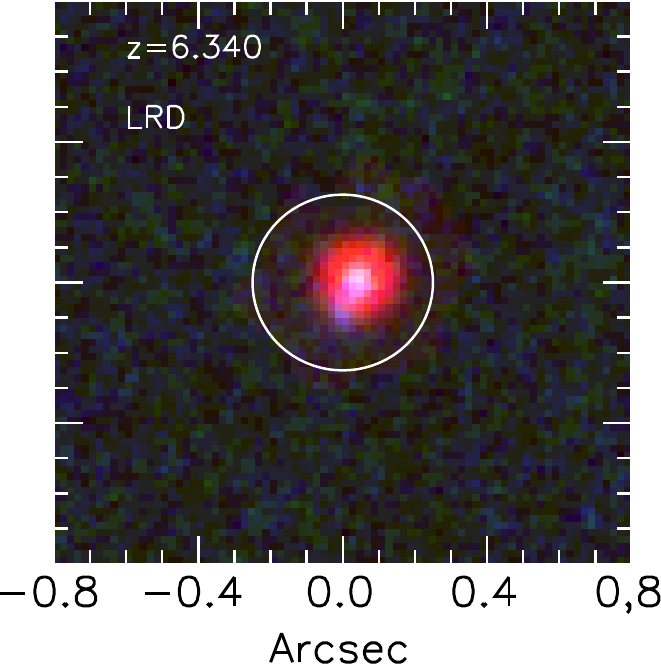}
\caption{
RGB images (F444W, F200W, and F150W) for six of the LRDs 
(DD20, DD14, DD58, DD34, DD16, and DD22)
with a neighboring blue companion. The
white circle shows a $0\farcs25$ radius.  
In all cases, the blue companion is within $0\farcs25$ of the LRD.
We show the redshift in the upper left corner. All of these sources
have spectroscopic redshifts and have been classified as LRDs in
the literature. \cite{golubchik26} found a similar pair behind
the cluster A383.
\label{fig:doubleblue}
}
\end{figure*}

Up to this point, we have considered only the sizes and
separations of the sources. However, it is clear that
adding the pairs criterion significantly increases the
fractional LRD selection. Thus, we have 11 published LRDs out
of 31 sources in the compact pairs sample versus 24 out of
153 in the full compact sample. 

More specifically, there are a large fraction of the
pairs with one blue and one red member. Only one pair
has two red members, while a number of pairs have two blue
members. For each member of a pair, we measured the fluxes
over the full available bands. 
We measured the counts in each component using the Gaussian fit
and converted this to the flux.
In Table~\ref{tab:fortex_photometry}, we give the fluxes of the pairs
in $\mu$Jy. Unlike the UNCOVER fluxes 
aperrure correction so, in comparing the cmnbined fluxes with the
uncover data, we need to correct for this offset.
(See Figure~8).
We then  used the SED for each individual 
member to determine if it is an LRD. 

Different papers have adopted 
different selection criteria for determining whether a source is an LRD. 
We use the power-law fitting method of \citet{kocevski25},
which compares power law fits to the 
 shapes of the rest-frame UV and optical portions 
of the SED to determine whether the source has the v-shape that is
characteristic of an LRD. V-shaped sources have power law
indices $\beta$
 (f$_\lambda \propto \lambda^\beta$) less than -0.37 in the rest frame UV and larger than 0 in the 
rest frame optical.
We note that separating the pairs can make it easier for one member to satisfy 
this criterion. We find that 14 of the pairs have at least one member that does 
(labeled ``v-shape" in the last column of Table~\ref{tab:full_data}),
11 of which are already included in the known LRD sample. 

The pair DD54 and DD55 is unique in containing two red members,
one of which individually satisfies the v-shape criterion. In Figure~\ref{fig:doublered},
we show the image of this pair.
The more common pairing is an LRD with a blue companion
(see Figure~\ref{fig:doubleblue}), or a pair of blue sources.

\subsection {Spectra}
\label{sec:redshifts}
We next searched the JWST DAWN archive (\citealt{heintz25}) for sources 
with NIRSpec slit spectra
whose orientation is such that the spectra overlay both 
members of the pair and where the separation
is large enough that we can distinguish the two sources. 
We used the overlay program in the DAWN archive (\citealt{hausen22})
to find the spectra that were suitable for this purpose.
We found useful spectra for the pairs DD16 and DD17; DD34 and DD35;
DD54 and DD55; and DD58 and DD59. We then downloaded the reduced 
NIRSpec spectra for these sources from the DAWN archive \citep{degraaff25}. 
We did not attempt to use WFSS spectra.

\begin{figure*}[t]
\includegraphics[width=3.5in,angle=0]{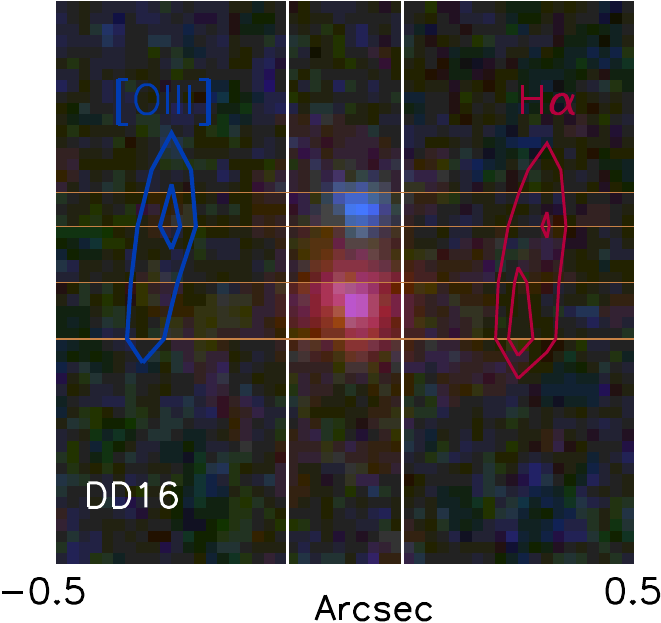}
\hskip 0.5cm
\includegraphics[width=3.5in,angle=0]{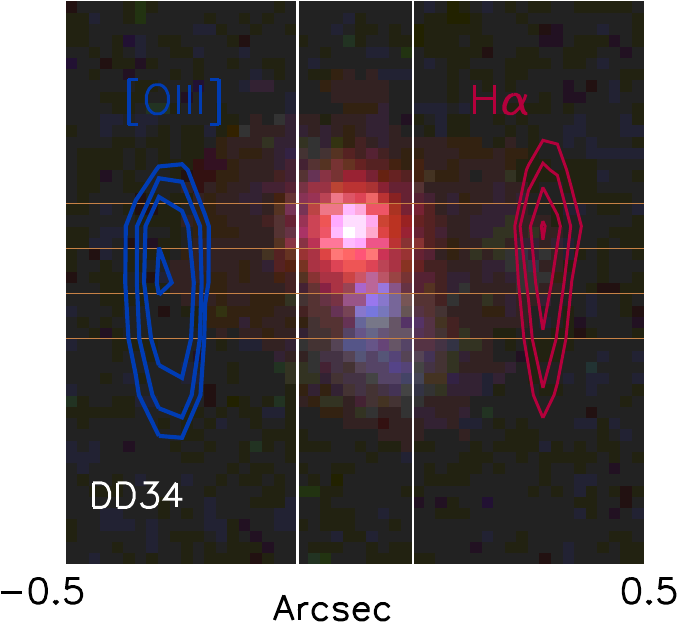}
\caption{RGB images (F444W, F200W, and F150W) for two of the pairs 
where the slit for the NIRSpec spectrum
was roughly aligned to cover both sources in the pair (slit position marked
by vertical white lines): (left panel) DD16 and DD17 and (right panel) DD34 and DD35.
Both of these pairs have fairly wide separations ($0\farcs165$ and $0\farcs149$,
respectively), making it possible to see the spatial structure in the emission
lines. In each panel, we show a contour plot of the (left) [OIII]5007
(blue) and (right) H$\alpha$ (red) emission lines.
The horizontal gold lines mark the positions of the two members of the pairs.
For both pairs, the emission lines extend over both sources, and we can see 
a small velocity offset between the two. 
\label{fig:Ha_contour}
}
\end{figure*}

\begin{figure}
\vskip 0.25cm
\includegraphics[width=3.25in,angle=0]{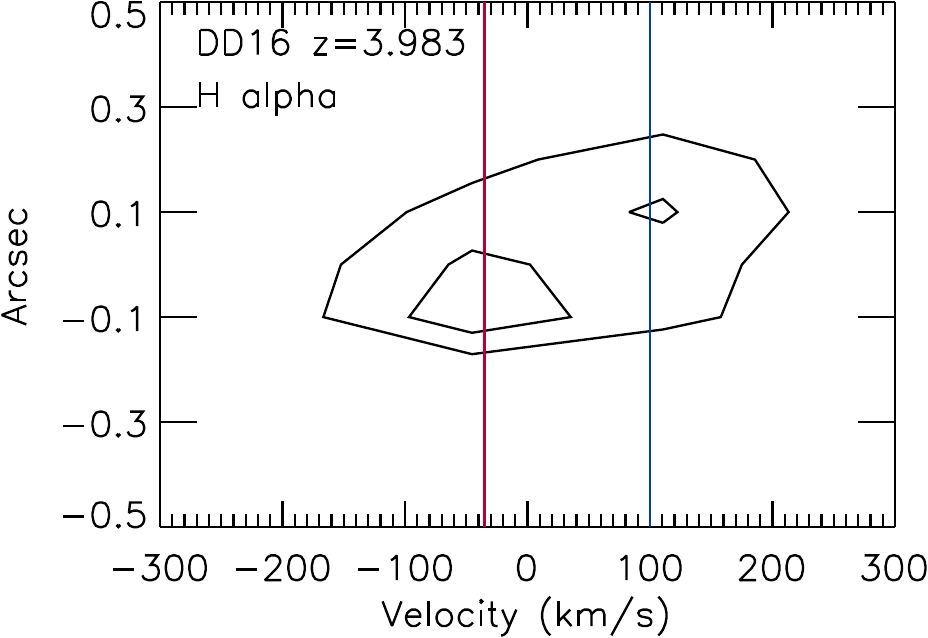}
\caption{
Contour plot of the 2D spectral image of the H$\alpha$ emission line
in the DD16 and DD17 pair. The red and blue vertical lines
show, respectively, the velocities of the red (DD16) and blue (DD17) sources, 
with a velocity separation of 145~km~s$^{-1}$.
\label{fig:spec_dd16}
}
\end{figure}

The spectra for the pair DD16 and DD17 are the clearest, both because of their wider 
separation (0\farcs165) and because of the high resolution and high S/N
G130M spectra from PID 09214 (PI C. Mason). 
In Figure~\ref{fig:Ha_contour} (left panel),
we show the NIRSpec slit (white lines) overlaid on the pair, illustrating
that both members are well covered by the slit. 
We also show contour plots of the [OIII]5007 (blue) and H$\alpha$ (red)
lines in the spectra. The $x$ position and scale is arbitrary,
but the $y$-axis is matched to the image, and the position of
the peaks in the spectra correspond to the object positions
(horizontal gold lines). The red source in the pair has stronger
H$\alpha$ emission, while the blue source has stronger [OIII]5007 emission. 
The two components show a velocity separation of 145~km~s$^{-1}$
(see Figure~\ref{fig:spec_dd16}). We discuss the spectra further below.

In Figure~\ref{fig:Ha_contour} (right panel), we show a similar
plot for the pair DD34 and DD35 from PID 02561 (PI Labbe).
Here, both sources lie within the slit, but
they are closer (0\farcs149) and more diffuse, resulting in
more blending in the spectral image. This PRISM spectrum
has lower resolution (R $\sim350$), which makes
the relative velocity structure in the two sources harder
to differentiate visually. However, the [OIII]5007 emission line is clearly
more weighted to the blue source than the H$\alpha$ emission line, which is
strongest in the red source.

There are two spectra for the pair DD54 and DD55:  a PRISM spectrum from 
PID 02561 (PI Labe) and a G395M spectrum from PID 06053 (PI Wold).
The Wold spectrum is of much higher quality, but it is slightly
misaligned with the slit, resulting in the lines from
the two sources being blended. The Labbe spectrum 
shows the H$\alpha$ emission line with a peak at the position of the 
reddest source and the [OIII]5007 emission line displaced towards the
second member of the pair. The pair DD58 and DD59 has only
a PRISM spectrum from Labbe. In this case, it is unclear
whether the blue member of the pair is present in the emission lines.

\section{Discussion}
\label{discuss}
\subsection{SEDs}
\label{seds}
The most direct effect of the doubles is that we may fail
to determine properly the SEDs of the sources, and, in particular,
to identify an LRD due to contamination by its neighbor.
The composite SEDs contain contributions
from both members of the pair, and the blue members
may contribute a substantial part of the rest-frame UV flux;
in some cases, they may dominate at these wavelengths
(see Table~\ref{tab:fortex_photometry}). Pairs of 
this type include DD20,DD21, DD30,DD31, DD32,DD33, and DD58,DD59.

\begin{figure}[t]
\includegraphics[width=3.4in,angle=0]{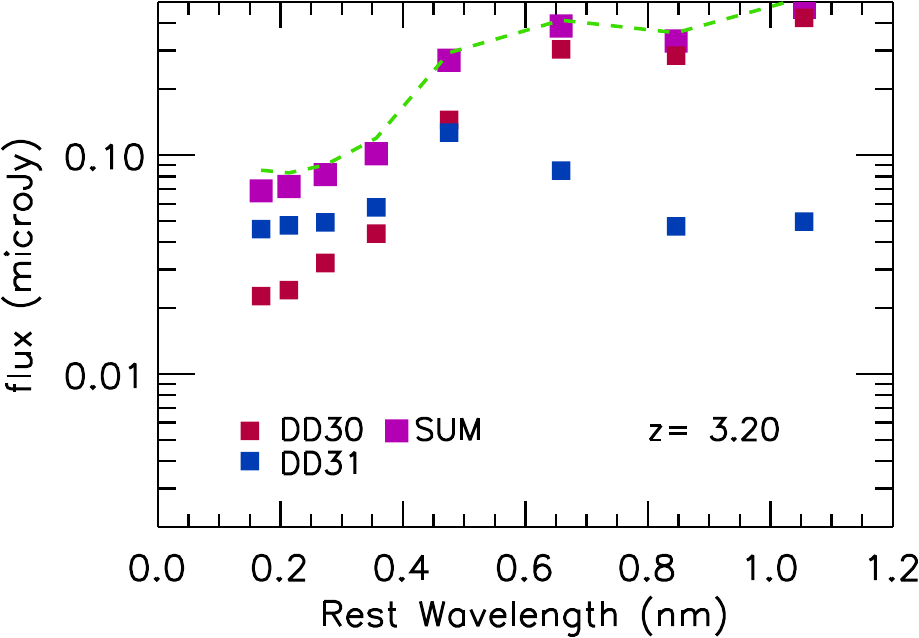}
\caption{Relative SEDs of the DD30 (red squares) and DD31 (blue squares) pair. 
The combined fluxes are shown
in purple and closely match  the UNCOVER catalog fluxes
when the aperture correction used for the UNCOVER fluxes is  removed (green dashed curve).
DD31 dominates the fluxes in the rest-frame UV, and DD30 in the 
rest-frame optical.
\label{fig:flux_pair}
}
\end{figure}

In Figure~\ref{fig:flux_pair}, we show an example of this
effect for the DD30 and DD31 pair.
Here, the composite SED (purple squares) marginally fails to meet the v-shape
criterion for LRD selection, having too flat a rest-frame optical SED.
However, DD30 (red squares) does meet the v-shape criterion.
DD31 (blue squares) appears to be a relatively
unobscured star-forming galaxy based on its flat SED.

For all 11 pairs in the sample whose composite SEDs were
previously classified as LRDs, we find one member's SED individually 
meets the v-shape criterion. However, in a further three pairs, including the
DD30 and DD31 pair shown in Figure~\ref{fig:flux_pair}, we find that one 
member's SED also individually meets the v-shape criterion, but that fact 
was previously hidden due to the contribution from the companion.

\subsection{2D Spectra}
\label{specs}
An interesting question to ask is, from which member of the pair do
broad lines arise?  
For example, the \citet{harikane23} sources, which correspond to the DD2 and DD3  
and DD4 and DD5 pairs, are BLAGNs, so we may investigate which member hosts 
the AGN when the broad line is seen in the 2D spectra. 

For the DD4 and DD5 pair, we have two spectra. The Treu
spectrum (PID 01324), which was used by \citet{harikane23}, is
centered on the blue member and clearly
shows the broad line in H$\alpha$. A later spectrum from Mason
(Program ID 09214) is centered on the red member
and does not clearly show the broad line in H$\alpha$.
We conclude that, in this case, it is the blue member
that hosts the AGN.

\begin{figure}[t]
\center{
\includegraphics[width=2.5in,angle=0]{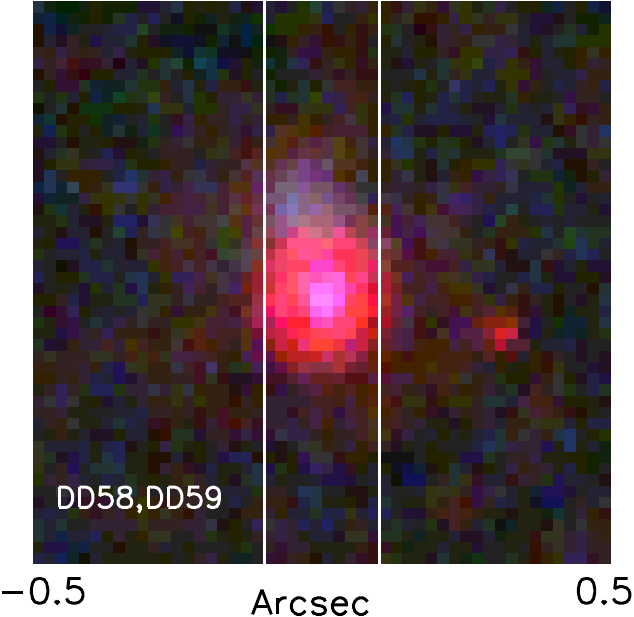}
\vskip 0.15cm
\includegraphics[width=2.95in,angle=0]{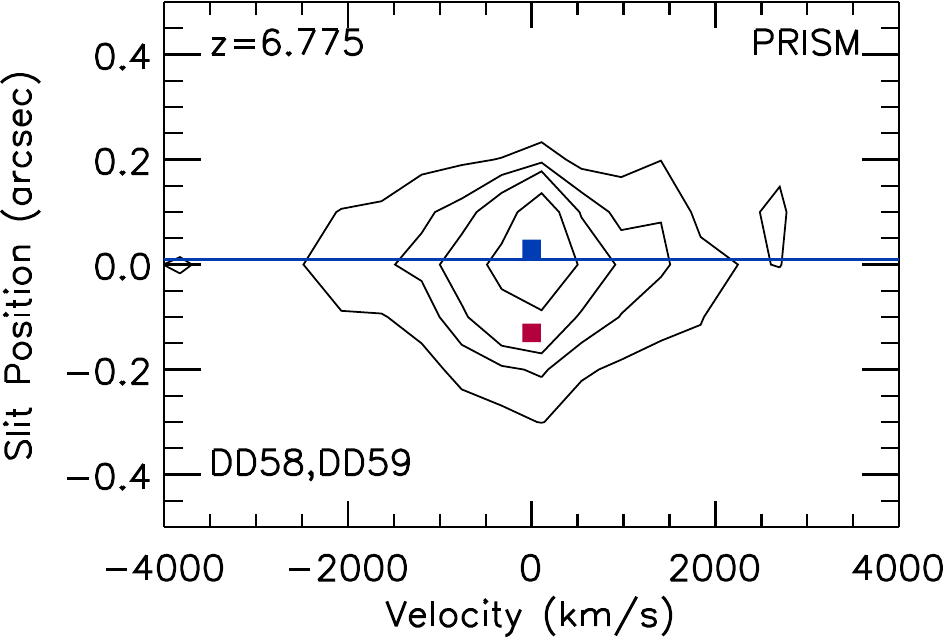}
\vskip 0.15cm
\includegraphics[width=3.15in,angle=0]{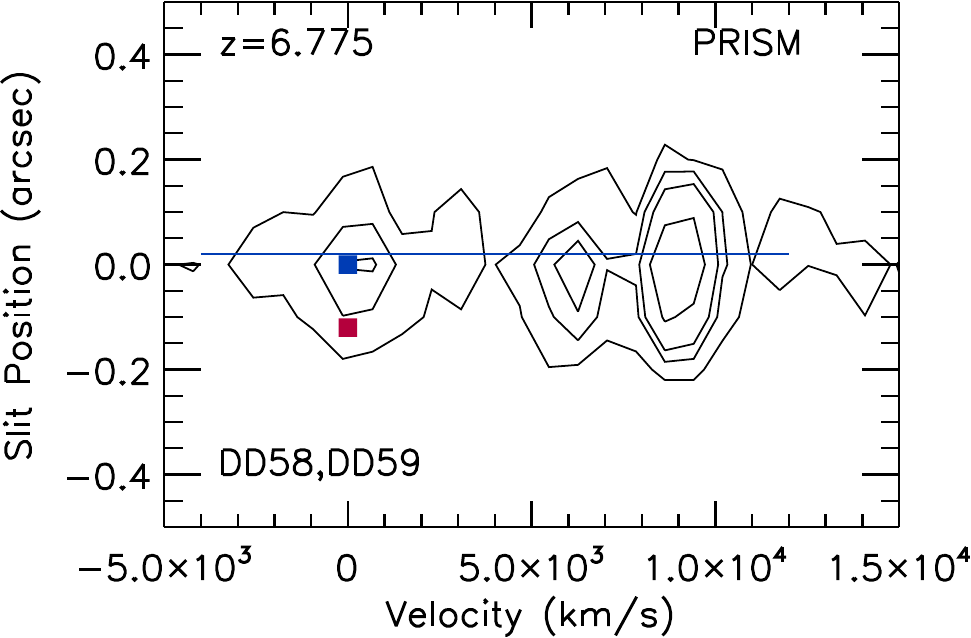}
\caption{2D spectra of the pair DD58 and DD59.
(Top) Orientation along the slit.
DD58 is the lower, redder object, and DD59 is the upper, bluer
object. 
(Middle) 2D spectrum of H$\alpha$.
(Bottom) 2D spectrum of the H$\beta$ and [OIII] region.
The broad line is easily visible and marked in each panel
with the horizontal blue line. The relative source positions
are shown with the blue (DD59) and red (DD58) squares.
The broad lines are clearly associated with the blue companion
DD59 and not with the LRD DD58.
\label{fig:2d_58}
}
}
\end{figure}

In Figure~\ref{fig:2d_58}, we show the DD58 and DD59 pair.
Here, the broad line can be clearly seen in H$\alpha$ (center panel),
and, more marginally, in H$\beta$ (right panel).
For both emission lines, the broad line appears to arise
in the northerly member of the pair (DD59; see slit configuration in left panel).
The LRD (DD58) seems to contain only narrow lines.

\begin{figure}[t]
\center{
\includegraphics[width=2.5in,angle=0]{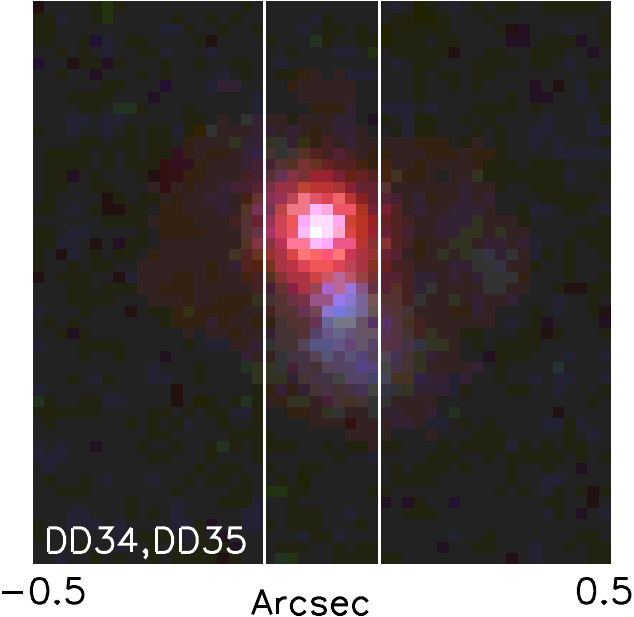}
\vskip 0.15cm
\includegraphics[width=2.95in,angle=0]{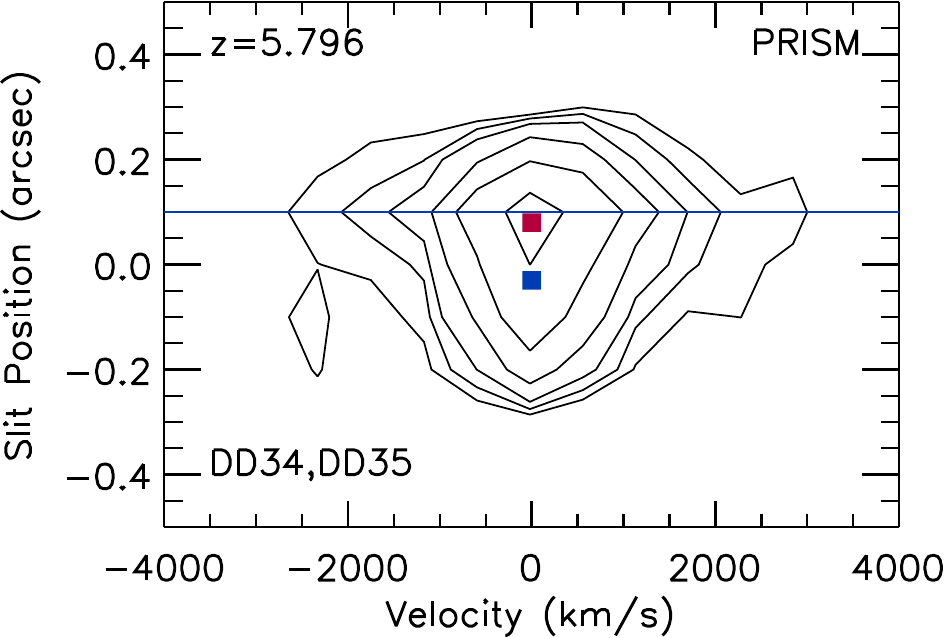}
\caption{2D spectra of the pair DD34 and DD35.
(Top) Orientation along the slit. DD34 is the upper, redder object, 
and DD35 is the lower, bluer object. 
(Bottom) 2D spectrum of H$\alpha$.  
The broad line is easily visible and marked
with the horizontal blue line. 
The relative source positions are shown with the blue (DD35) and red (DD34) squares.
The broad line is clearly associated with the LRD DD34.
\label{fig:2d_34}
}
}
\end{figure}

In Figure~\ref{fig:2d_34}, we show the DD34 and DD35 pair. Here,
the broad line appears to coincide instead with the LRD (i.e., DD34; 
upper red source). The blue companion (DD35) has narrow emission lines.

In conclusion, the broad line can arise in either component
of the pair. It does not necessarily arise in an LRD
member or in the redder member of a pair.

\section{Interpretation}
\begin{figure}
\includegraphics[width=3.2in,angle=0]{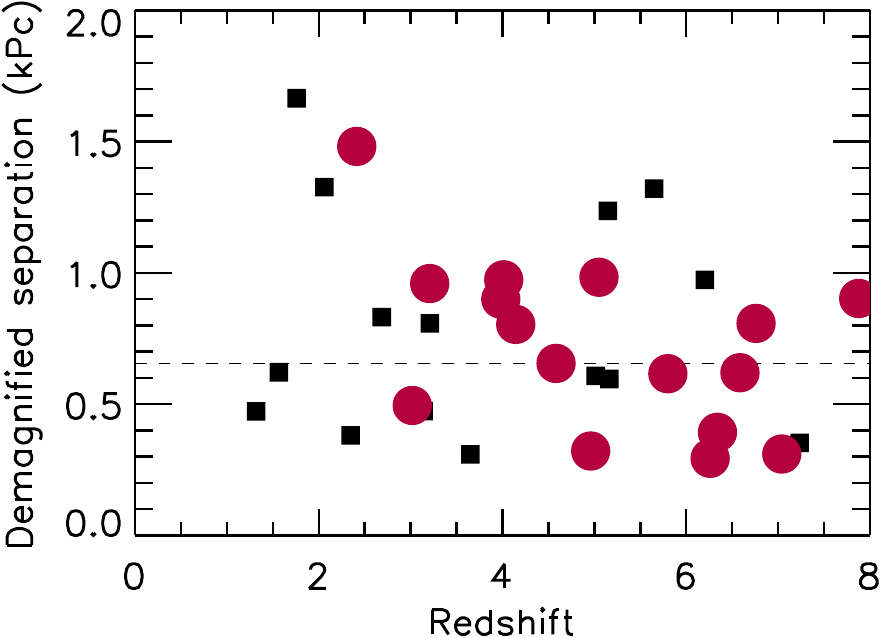}
\caption{Physical separation of the pairs vs.
redshift. The separations are corrected for lensing (see text).
The red circles highlight pairs where at least one member has broad lines 
or a v shape. The horizontal dashed line shows the median separation of 
the sample.
\label{fig:show_separations}
}
\end{figure}

We have demonstrated that substructure seems to be a common 
feature of the broad-line systems at high redshift, and, in particular, of the LRDs.
In Figure~\ref{fig:show_separations}, we show the demagnified physical separations 
of the pairs vs. redshift. We highlight with red circles pairs where at least one member 
has broad lines or a v-shape. 
Separations range from 0.3 to 1.7~kpc and show little dependence on 
redshift. The range of separations appears similar for both populations,
for which there are a number of possible explanations.

The most straightforward interpretation may be
that the substructure represents galaxy mergers
triggering AGN activity and shrouding
the system in gas and dust. This would naturally explain why close pairs 
are such strong markers of BLAGNs and could also
explain many of the unusual properties of LRDs.

However, there might also have been enhancement of 
supermassive black hole growth if a DCBH formed
due to the presence of a companion, 
which allowed early collapse of a seed black hole by suppressing molecular H cooling
\citep{baggen26}. This could have driven the growth of the black hole in one of the pair.
Then, as the pair begins to merge, 
it could trigger a super-Eddington accretion event shrouded in dust.

However, we should also bear in mind the possibility 
that these are not AGNs at all and that the substructure
reflects the formation of globular clusters, as in the \citet{chisolm26} model.

Here we focus on the accretion model. We first computed the stellar masses of each of the 
members using the rest-frame demagnified $J$-band luminosity.
For the higher redshift sources, this involves a substantial extrapolation. The highest value
is a logarithmic mass of 9.5~M$_\odot$, and the median value
is 6.9~M$_\odot$. Given the possible contributions of the AGNs,
these may be best viewed as upper limits. Given these uncertainties,
a more sophisticated SED fitting calculation does not seem justified.

We can determine whether the pairs might be merging based on their masses, physical separations,
and velocity separations. We demagnified all quantities using the mean pair lensing 
magnification from the UNCOVER lensing model \citep{furtak23}. 
In Table~\ref{tab:mergers}, we summarize the 
magnifications ($\mu$), the adopted 
redshifts ($z_{spec}$, where available, and otherwise $z_{phot}$, as
given in Table~1), and
observed separations ($\theta$), together with the delensed physical
separations ($r$) and the delensed stellar masses of the two members 
($r=r_\mathrm{obs}/\sqrt{\mu}$ and $M_\star=M_\star^\mathrm{obs}/\mu$; the latter
are labeled as either ``A" or ``B" in the order of the listed IDs of the members).  

We calculate total masses ($M_{\rm tot}$) assuming a dark matter ratio of 30, and 
we combine these masses with the physical separations to compute escape velocities ($v_{\rm esc}$). 

We also provide the mass ratio ($q$) and the merger type based
on the mass ratio: major ($q>0.25$), minor ($0.10<q\le0.25$), and satellite ($q\le0.10$).

\begin{figure}[t]
\includegraphics[width=3.2in,angle=0]{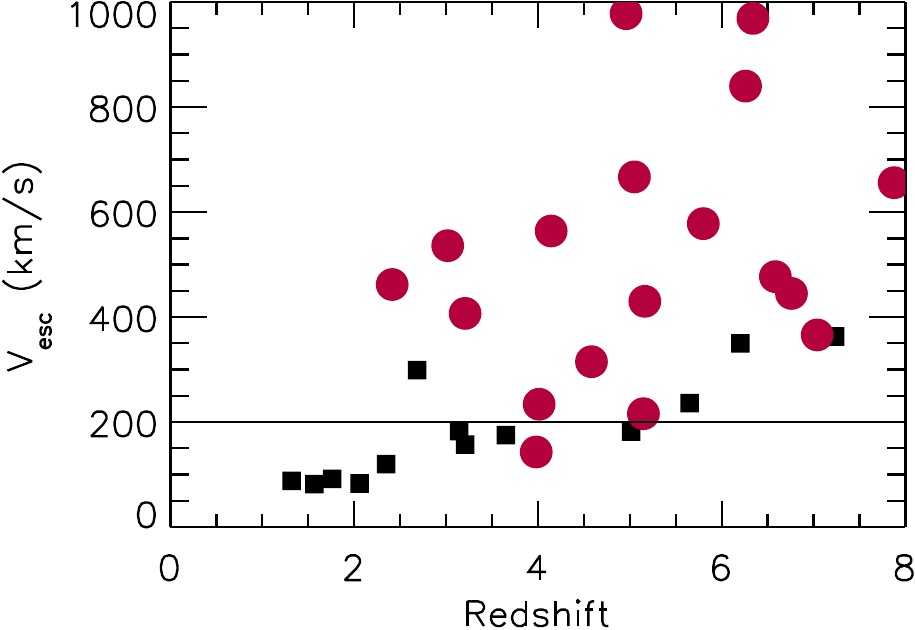}
\caption{Escape velocity of the pairs vs. redshift.
The red circles highlight pairs where at least one member has broad lines 
or a v shape. The horizontal 
line shows a velocity of 200~km~s$^{-1}$. 
17 of the 21 sources lying above this are either BLAGNs 
(BLAGNs are labeled as such in the Class 
column of Table~\ref{tab:mergers}) or have v shapes
(v shapes are marked by daggers in the ID column of Table~\ref{tab:mergers}). 
Only one of the 10 more weakly bound pairs does.
\label{fig:vesc_z}
}   
\end{figure}

In Figure~\ref{fig:vesc_z}, we show the escape velocity of the pairs versus redshift.
We highlight with red circles sources where at least one member has broad lines or a v shape.
The results appear strongly consistent with the accretion hypothesis.
The red circles sources are much more tightly bound.
Of the 21 pairs with escape velocities $>200$~km~s$^{-1}$, 17 have LRD or BLAGN 
signatures. Meanwhile, only one (DD16 and DD17) of the 10 more weakly bound pairs does. 
In this case, the small velocity separation in the spectra would also be consistent
with the pair being bound.

\section{Summary}
\label{sec:summary}
We used deep JWST imaging of the lensing cluster field A2744 to search 
systematically for close pairs among compact sources. We identified 31 ``Double Dot'' 
pairs with separations $<0\farcs25$ and a median separation of $0\farcs15$. 
We compared these with sources in the field that were previously classified as 
LRDs or BLAGNs. Our main findings are as follows:

\begin{enumerate}

\item At least 16 of the 24 previously published LRDs in the A2744 field ($\sim67$\%) 
are members of close pairs, as are both of the high-redshift BLAGNs identified by 
\citet{harikane23}. Since most LRDs contain broad emission lines, close pairs are 
therefore extremely effective markers of galaxies with broad lines at high redshift.

\item We found that companion contamination can significantly distort SED measurements. 
The blue companion in a pair may dominate  the rest-frame UV flux, which can mask 
the ``v-shape'' signature used to classify LRDs. When we decomposed the pairs and 
analyzed each member individually, we found three additional sources that 
satisfied the v-shape criterion for an LRD.

\item We performed 2D spectroscopic analysis of four pairs, finding that broad 
emission lines can arise in either member of the pair; they are not confined to the
redder or LRD component. For example, in the pair DD58 and DD59, broad lines
originate in the blue companion DD59, while in the pair DD34 and DD35,
in the LRD DD34.

\item We determined that sources with escape velocities $>200$~km~s$^{-1}$
overwhelmingly showed broad lines or a v shape (17 of 21), while in 
the more weakly bound pairs, only 1 of 10 did.
This is strongly consistent with a merger-driven accretion scenario.

\end{enumerate}

\clearpage
\startlongtable
\begin{deluxetable*}{rcccccccc}
\renewcommand\baselinestretch{1.0}
\tablecaption{Compact Pairs Sample \label{tab:full_data}}
\scriptsize
\tablehead{ID & Offset & R.A. & Decl. & FWHM & $z_{phot}$ & $z_{spec}$ & Class & v shape \\
 & (arcsec) & (deg) & (deg) & (arcsec) & & & & \\
 (1) & (2) & (3) & (4) & (5) & (6) & (7) & (8) & (9)
 }
 \startdata
0 & 0.118 & 3.6326168 & $-30.436611$ & 0.079 & 2.686 & $\cdots$ & $\cdots$ & $\cdots$ \\
1 & $\cdots$ & 3.6325788 & $-30.436666$ & 0.188 & $\cdots$ & $\cdots$ & $\cdots$ & $\cdots$ \\
2 & 0.178 & 3.5803041 & $-30.424389$ & 0.073 & $\cdots$ & 4.013 & BLAGN & $\cdots$ \\
3 & $\cdots$ & 3.5802500 & $-30.424389$ & 0.169 & $\cdots$ & $\cdots$ & $\cdots$ & $\cdots$ \\
4 & 0.127 & 3.5771418 & $-30.422527$  & 0.083 & $\cdots$ & 4.583 & BLAGN & $\cdots$ \\
5 & $\cdots$ & 3.5771585 & $-30.422556$ & 0.115 & $\cdots$ & $\cdots$ & $\cdots$ & $\cdots$ \\
6 & 0.133 & 3.6081877 & $-30.422001$ & 0.070 & $\cdots$ & 5.015 & $\cdots$ & $\cdots$ \\
7 & $\cdots$ & 3.6082001 & $-30.421972$ & 0.073 & $\cdots$ & $\cdots$ & $\cdots$ & $\cdots$ \\
8 & 0.081 & 3.5537708 & $-30.410110$ & 0.075 & 7.235 & $\cdots$ & $\cdots$ & $\cdots$ \\
9 & $\cdots$ & 3.5537875 & $-30.410110$ & 0.068 & $\cdots$ & $\cdots$ & $\cdots$ & $\cdots$ \\
10 & 0.077 & 3.5789330 & $-30.408138$ & 0.065 & 2.347 & $\cdots$ & $\cdots$ & $\cdots$ \\
11 & 
$\cdots$ & 3.5789042 & $-30.408138$ & 0.103 & $\cdots$ & $\cdots$ & $\cdots$ & $\cdots$ \\
12 & 0.068 & 3.5775459 & $-30.407444$ & 0.072 & 3.650 & $\cdots$ & $\cdots$ & $\cdots$ \\
13 & $\cdots$ & 3.5775540 & $-30.407473$ & 0.150 & $\cdots$ & $\cdots$ & $\cdots$ & $\cdots$ \\
14 & 0.211 & 3.5567081 & $-30.408167$ & 0.068 & $\cdots$ & 2.414 & LRD & yes \\
15 & $\cdots$ & 3.5567374 & $-30.408138$ & 0.069 & $\cdots$ & $\cdots$ & $\cdots$ & $\cdots$ \\
16 & 0.165 & 3.5677586 & $-30.407278$ & 0.076 & $\cdots$ & 3.982 & LRD & yes \\
17 & $\cdots$ & 3.5677626 & $-30.407251$ & 0.075 & $\cdots$ & $\cdots$ & $\cdots$ & $\cdots$ \\
18 & 0.184 & 3.5508418 & $-30.406584$ & 0.070 & $\cdots$ & 5.050 & LRD & yes \\
19 & $\cdots$ & 3.5507791 & $-30.406610$ & 0.184 & $\cdots$ & $\cdots$ & $\cdots$ & $\cdots$ \\
20 & 0.152 & 3.6125040 & $-30.400583$ & 0.083 & $\cdots$ & 6.585 & LRD & yes \\
21 & $\cdots$ & 3.6125374 & $-30.400555$ & 0.083 & $\cdots$ & $\cdots$ & $\cdots$ & $\cdots$ \\
22 & 0.087 & 3.6206081 & $-30.399944$ & 0.076 & $\cdots$ & 6.340 & LRD & yes \\
23 & $\cdots$ & 3.6206207 & $-30.399971$ & 0.095 & $\cdots$ & $\cdots$ & $\cdots$ & $\cdots$ \\
24 & 0.186 & 3.5330207 & $-30.386862$ & 0.112 & 2.060 & $\cdots$ & $\cdots$ & $\cdots$ \\
25 & $\cdots$ & 3.5329959 & $-30.386889$ & 0.177 & $\cdots$ & $\cdots$ & $\cdots$ & $\cdots$ \\
26 & 0.263 & 3.6236458 & $-30.384249$ & 0.091 & $\cdots$ & 5.651 & $\cdots$ & $\cdots$ \\
27 & $\cdots$ & 3.6236210 & $-30.384306$ & 0.077 & $\cdots$ & $\cdots$ & $\cdots$ & $\cdots$ \\
28 & 0.251 & 3.6138084 & $-30.382999$ & 0.082 & 5.148 & $\cdots$ & $\cdots$ & $\cdots$ \\
29 & $\cdots$ & 3.6138418 & $-30.382917$ & 0.115 & $\cdots$ & $\cdots$ & $\cdots$ & $\cdots$ \\
30 & 0.181 & 3.5967584 & $-30.382557$ & 0.082 & 3.208 & $\cdots$ & $\cdots$ & yes \\
31 & 
$\cdots$ & 3.5967040 & $-30.382557$ & 0.125 & $\cdots$ & $\cdots$ & $\cdots$ & $\cdots$ \\
32 & 0.065 & 3.6179543 & $-30.381834$ & 0.069 & 1.318 & $\cdots$ & $\cdots$ & $\cdots$ \\
33 & $\cdots$ & 3.6179709 & $-30.381805$ & 0.125 & $\cdots$ & $\cdots$ & $\cdots$ & $\cdots$ \\
34 & 0.149 & 3.5428374 & $-30.380638$ & 0.065 & $\cdots$ & 5.800 & LRD & yes \\
35 & $\cdots$ & 3.5427957 & $-30.380638$ & 0.161 & $\cdots$ & $\cdots$ & $\cdots$ & $\cdots$ \\
36 & 0.242 & 3.6045001 & $-30.380445$ & 0.096 & $\cdots$ & 7.877 & $\cdots$ & yes \\
37 & $\cdots$ & 3.6045625 & $-30.380388$ & 0.137 & $\cdots$ & $\cdots$ & $\cdots$ & $\cdots$ \\
38 & 0.226 & 3.5328708 & $-30.379862$ & 0.097 & $\cdots$ & 6.203 & $\cdots$ & $\cdots$ \\
39 & $\cdots$ & 3.5328500 & $-30.379917$ & 0.070 & $\cdots$ & $\cdots$ & $\cdots$ & $\cdots$ \\
40 & 0.122 & 3.6291165 & $-30.378445$ & 0.070 & 3.209 & $\cdots$ & $\cdots$ & $\cdots$ \\
41 & $\cdots$ & 3.6291039 & $-30.378416$ & 0.146 & $\cdots$ & $\cdots$ & $\cdots$ & $\cdots$ \\
42 & 0.228 & 3.5143626 & $-30.377140$ & 0.149 & 1.759 & $\cdots$ & $\cdots$ & $\cdots$ \\
43 & $\cdots$ & 3.5142877 & $-30.377167$ & 0.139 & $\cdots$ & $\cdots$ & $\cdots$ & $\cdots$ \\
44 & 0.100 & 3.5337749 & $-30.376446$ & 0.069 & 1.566 & $\cdots$ & $\cdots$ & $\cdots$ \\
45 & $\cdots$ & 3.5337584 & $-30.376446$ & 0.139 & $\cdots$ & $\cdots$ & $\cdots$ & $\cdots$ \\
46 & 0.149 & 3.5838125 & $-30.374500$ & 0.112 & 5.164 & $\cdots$ & $\cdots$ & $\cdots$ \\
47 & $\cdots$ & 3.5838542 & $-30.374527$ & 0.091 & $\cdots$ & $\cdots$ & $\cdots$ & $\cdots$ \\
48 & 0.080 & 3.5194333 & $-30.375139$ & 0.071 & 3.018 & $\cdots$ & $\cdots$ & yes \\
49 & $\cdots$ & 3.5194249 & $-30.375168$ & 0.080 & $\cdots$ & $\cdots$ & $\cdots$ & $\cdots$ \\
50 & 0.110 & 3.5696080 & $-30.373247$ & 0.086 & $\cdots$ & 7.040 & LRD & yes \\
51 & 
$\cdots$ & 3.5695920 & $-30.373220$ & 0.127 & $\cdots$ & $\cdots$ & $\cdots$ & $\cdots$ \\
52 & 0.082 & 3.5190916 & $-30.371307$ & 0.103 & 3.142 & $\cdots$ & $\cdots$ & $\cdots$ \\
53 & $\cdots$ & 3.5190876 & $-30.371279$ & 0.092 & $\cdots$ & $\cdots$ & $\cdots$ & $\cdots$ \\
54 & 0.091 & 3.5790000 & $-30.362583$ & 0.108 & $\cdots$ & 6.260 & LRD & yes \\
55 & $\cdots$ & 3.5789750 & $-30.362583$ & 0.093 & $\cdots$ & $\cdots$ & $\cdots$ & $\cdots$ \\
56 & 0.057 & 3.5299833 & $-30.358112$ & 0.071 & $\cdots$ & 4.963 & LRD & yes \\
57 & $\cdots$ & 3.5299666 & $-30.358112$ & 0.108 & $\cdots$ & $\cdots$ & $\cdots$ & $\cdots$ \\
58 & 0.172 & 3.5339832 & $-30.353306$ & 0.140 & $\cdots$ & 6.760 & LRD & yes \\
59 & $\cdots$ & 3.5340335 & $-30.353277$ & 0.115 & $\cdots$ & $\cdots$ & $\cdots$ & $\cdots$ \\
60 & 0.158 & 3.6102083 & $-30.421000$ & 0.082 & 4.144 & $\cdots$ & LRD & yes \\
61 & $\cdots$ & 3.6101875 & $-30.421028$ & 0.137 & $\cdots$ & $\cdots$ & $\cdots$ & $\cdots$ \\
\enddata
\tablecomments{
(1) ID, (2) offset between the two members of each pair rounded to three decimal places, 
(3) Right Ascension, (4) Declination, (5) FWHM rounded to three decimal places,
(6) photometric redshift, (7) spectroscopic redshift, (8) previously identified class (BLAGN or LRD), 
and (9) whether the member of the pair satisfies the v-shape criterion for an LRD from \citet{kocevski25}.
To maintain clarity for the pairs, we only list redshift values for the primary member of each pair 
(i.e., even IDs).
}
\end{deluxetable*}

\startlongtable
\begin{deluxetable*}{rccccccccc}
\renewcommand\baselinestretch{1.0}
\tablecaption{Photometry for the Compact Pairs Sample \label{tab:fortex_photometry}}
\scriptsize
\tablehead{ID & $z$ & F70W & F90W & F115W & F150W & F200W & F277W & F356W & F444W \\
 & & ($\mu$Jy) & ($\mu$Jy) & ($\mu$Jy) & ($\mu$Jy) & ($\mu$Jy) & ($\mu$Jy) & ($\mu$Jy) & ($\mu$Jy) \\ 
 (1) & (2) & (3) & (4) & (5) & (6) & (7) & (8) & (9) & (10)
 }
\startdata
0 & 2.686 & 0.000 & 0.000 & 0.069 & 0.088 & 0.177 & 0.109 & 0.098 & 0.154 \\
1 & 2.686 & 0.000 & 0.000 & 0.046 & 0.048 & 0.082 & 0.033 & 0.035 & 0.037 \\
2 & 4.013 & 0.024 & 0.036 & 0.041 & 0.054 & 0.076 & 0.196 & 0.140 & 0.077 \\
3 & 4.013 & 0.022 & 0.024 & 0.018 & 0.021 & 0.026 & 0.042 & 0.028 & 0.013 \\
4 & 4.583 & 0.019 & 0.033 & 0.037 & 0.036 & 0.048 & 0.118 & 0.100 & 0.083 \\
5 & 4.583 & 0.012 & 0.020 & 0.022 & 0.022 & 0.023 & 0.026 & 0.021 & 0.006 \\
6 & 5.015 & 0.011 & 0.015 & 0.014 & 0.013 & 0.017 & 0.034 & 0.020 & 0.025 \\
7 & 5.015 & 0.007 & 0.004 & 0.003 & 0.006 & 0.002 & 0.000 & 0.001 & 0.003 \\
8 & 7.235 & -0.001 & 0.000 & 0.009 & 0.010 & 0.007 & 0.011 & 0.011 & 0.020 \\
9 & 7.235 & 0.001 & 0.001 & 0.006 & 0.006 & 0.008 & 0.004 & 0.003 & 0.008 \\
10 & 2.347 & 0.031 & 0.029 & 0.029 & 0.032 & 0.032 & 0.032 & 0.031 & 0.030 \\
11 & 2.347 & 0.023 & 0.021 & 0.016 & 0.029 & 0.020 & 0.010 & 0.009 & 0.008 \\
12 & 3.650 & 0.027 & 0.027 & 0.023 & 0.022 & 0.028 & 0.037 & 0.016 & 0.019 \\
13 & 3.650 & 0.015 & 0.014 & 0.016 & 0.020 & 0.014 & 0.020 & 0.009 & 0.010 \\
14 & 2.414 & 0.016 & 0.018 & 0.023 & 0.046 & 0.152 & 0.504 & 0.849 & 1.027 \\
15 & 2.414 & 0.013 & 0.011 & 0.009 & 0.011 & 0.011 & 0.007 & 0.010 & 0.004 \\
16 & 3.982 & 0.008 & 0.008 & 0.009 & 0.008 & 0.010 & 0.015 & 0.010 & 0.003 \\
17 & 3.982 & 0.004 & 0.007 & 0.004 & 0.006 & 0.007 & 0.012 & 0.023 & 0.029 \\
18 & 5.050 & 0.008 & 0.015 & 0.022 & 0.027 & 0.035 & 0.155 & 0.272 & 0.412 \\
19 & 5.050 & 0.007 & 0.012 & 0.007 & 0.016 & 0.011 & 0.022 & 0.021 & 0.028 \\
20 & 6.585 & 0.001 & 0.006 & 0.019 & 0.020 & 0.023 & 0.028 & 0.067 & 0.102 \\
21 & 6.585 & 0.000 & 0.008 & 0.021 & 0.021 & 0.020 & 0.022 & 0.039 & 0.021 \\
22 & 6.340 & 0.001 & 0.002 & 0.009 & 0.013 & 0.017 & 0.027 & 0.113 & 0.291 \\
23 & 6.340 & 0.002 & 0.003 & 0.007 & 0.006 & 0.003 & 0.003 & -0.007 & -0.048 \\
24 & 2.060 & 0.014 & 0.010 & 0.018 & 0.025 & 0.028 & 0.031 & 0.030 & 0.028 \\
25 & 2.060 & 0.012 & 0.021 & 0.020 & 0.025 & 0.026 & 0.021 & 0.015 & 0.012 \\
26 & 5.651 & 0.004 & 0.022 & 0.034 & 0.034 & 0.035 & 0.045 & 0.096 & 0.057 \\
27 & 5.651 & -0.003 & 0.004 & 0.006 & 0.007 & 0.004 & 0.005 & 0.007 & 0.006 \\
28 & 5.148 & 0.013 & 0.028 & 0.030 & 0.027 & 0.028 & 0.071 & 0.027 & 0.048 \\
29 & 5.148 & 0.005 & 0.014 & 0.011 & 0.013 & 0.012 & 0.026 & 0.008 & 0.017 \\
30 & 3.208 & 0.026 & 0.026 & 0.033 & 0.045 & 0.149 & 0.307 & 0.286 & 0.423 \\
31 & 3.208 & 0.046 & 0.047 & 0.050 & 0.058 & 0.128 & 0.092 & 0.054 & 0.060 \\
32 & 1.318 & 0.000 & 0.007 & 0.014 & 0.026 & 0.025 & 0.022 & 0.017 & 0.029 \\
33 & 1.318 & 0.012 & 0.012 & 0.017 & 0.018 & 0.015 & 0.013 & 0.007 & 0.007 \\
34 & 5.800 & 0.002 & 0.009 & 0.019 & 0.025 & 0.035 & 0.050 & 0.138 & 0.156 \\
35 & 5.800 & 0.001 & 0.028 & 0.048 & 0.047 & 0.051 & 0.063 & 0.119 & 0.087 \\
36 & 7.877 & -0.003 & 0.000 & 0.048 & 0.082 & 0.076 & 0.083 & 0.094 & 0.227 \\
37 & 7.877 & 0.001 & 0.000 & 0.010 & 0.019 & 0.019 & 0.018 & 0.018 & 0.026 \\
38 & 6.203 & 0.001 & 0.020 & 0.039 & 0.040 & 0.041 & 0.057 & 0.141 & 0.104 \\
39 & 6.203 & 0.001 & 0.001 & 0.003 & 0.003 & 0.005 & 0.003 & 0.015 & 0.005 \\
40 & 3.209 & 0.014 & 0.014 & 0.016 & 0.023 & 0.060 & 0.042 & 0.022 & 0.032 \\
41 & 3.209 & 0.004 & 0.009 & 0.009 & 0.011 & 0.019 & 0.009 & 0.008 & 0.007 \\
42 & 1.759 & 0.032 & 0.034 & 0.057 & 0.077 & 0.070 & 0.062 & 0.055 & 0.051 \\
43 & 1.759 & 0.020 & 0.027 & 0.038 & 0.054 & 0.046 & 0.036 & 0.032 & 0.030 \\
44 & 1.566 & 0.032 & 0.039 & 0.063 & 0.023 & 0.023 & 0.040 & 0.024 & 0.031 \\
45 & 1.566 & 0.024 & 0.023 & 0.032 & 0.017 & 0.017 & 0.010 & 0.009 & 0.010 \\
46 & 5.164 & 0.012 & 0.037 & 0.043 & 0.044 & 0.050 & 0.256 & 0.076 & 0.185 \\
47 & 5.164 & 0.014 & 0.036 & 0.046 & 0.035 & 0.032 & 0.002 & 0.025 & 0.002 \\
48 & 3.018 & 0.020 & 0.018 & 0.018 & 0.015 & 0.031 & -0.013 & -0.022 & -0.036 \\
49 & 3.018 & 0.024 & 0.024 & 0.027 & 0.038 & 0.068 & 0.205 & 0.221 & 0.365 \\
50 & 7.040 & -0.001 & 0.000 & 0.007 & 0.004 & 0.005 & 0.005 & -0.001 & -0.003 \\
51 & 7.040 & 0.000 & 0.000 & 0.004 & 0.004 & 0.005 & 0.007 & 0.032 & 0.055 \\
52 & 3.142 & 0.002 & 0.005 & 0.009 & 0.018 & 0.043 & 0.051 & 0.035 & 0.042 \\
53 & 3.142 & 0.000 & 0.000 & 0.000 & 0.001 & -0.002 & -0.005 & -0.005 & -0.004 \\
54 & 6.260 & 0.001 & 0.033 & 0.078 & 0.073 & 0.074 & 0.103 & 0.248 & 0.227 \\
55 & 6.260 & 0.001 & 0.011 & 0.031 & 0.038 & 0.038 & 0.055 & 0.160 & 0.116 \\
56 & 4.960 & 0.010 & 0.024 & 0.023 & 0.025 & 0.027 & 0.092 & 0.238 & 0.291 \\
57 & 4.960 & 0.002 & 0.003 & 0.005 & 0.014 & 0.011 & 0.023 & 0.022 & 0.029 \\
58 & 6.760 & 0.003 & 0.004 & 0.011 & 0.014 & 0.010 & 0.017 & 0.054 & 0.085 \\
59 & 6.760 & 0.010 & 0.008 & 0.009 & 0.013 & 0.025 & 0.015 & 0.013 & 0.013 \\
60 & 4.144 & 0.062 & 0.073 & 0.076 & 0.088 & 0.108 & 0.300 & 0.449 & 0.424 \\
61 & 4.144 & 0.017 & 0.024 & 0.022 & 0.017 & 0.031 & 0.058 & 0.044 & 0.037 \\
\enddata
\tablecomments{
(1) ID, (2) redshift ($z_{spec}$, where available, and otherwise $z_{phot}$,
from Table~1), and JWST 
photometry from filters (3) F70W, (4) F90W, (5) F115W,
(6) F150W, (7) F200W, (8) F277W, (9) F356W, and (10) F444W.
}
\end{deluxetable*}

\begin{deluxetable}{lccccccccccc}
\tablecaption{Lensing-corrected Merger Analysis of Compact Pairs \label{tab:mergers}}
\tablecolumns{12}
\tablehead{
  \colhead{IDs} &
  \colhead{Class} &
  \colhead{$\mu$} &
  \colhead{$z$} &
  \colhead{$\theta$} &
  \colhead{$r$} &
  \colhead{$\log M_{\star,A}$} &
  \colhead{$\log M_{\star,B}$} &
  \colhead{$\log M_\mathrm{tot}$} &
  \colhead{$v_\mathrm{esc}$} &
  \colhead{$q$} &
  \colhead{Type} \\
  \colhead{} &
  \colhead{} &
  \colhead{} &
  \colhead{} &
  \colhead{(arcsec)} &
  \colhead{(kpc)} &
  \colhead{(M$_\odot$)} &
  \colhead{(M$_\odot$)} &
  \colhead{(M$_\odot$)} &
  \colhead{(km\,s$^{-1}$)} &
  \colhead{} &
  \colhead{} \\
   \colhead{(1)} &
  \colhead{(2)} &
  \colhead{(3)} &
  \colhead{(4)} &
  \colhead{(5)} &
  \colhead{(6)} &
  \colhead{(7)} &
  \colhead{(8)} &
  \colhead{(9)} &
  \colhead{(10} &
  \colhead{(11)} &
  \colhead{(12)}
}
\startdata
  DD0, DD1            & $\cdots$ & 1.33 & 2.68556 & 0.12 & 0.83 & 8.35 & 7.73 &  9.94 & 299 & 0.240 & Minor \\
  DD2, DD3            & BLAGN    & 1.68 & 4.01340 & 0.18 & 0.98 & 8.24 & 7.46 &  9.80 & 234 & 0.169 & Minor \\
  DD4, DD5            & BLAGN    & 1.68 & 4.58294 & 0.13 & 0.66 & 8.36 & 7.22 &  9.88 & 315 & 0.072 & Satellite \\
  DD6, DD7            & $\cdots$ & 1.96 & 5.01506 & 0.13 & 0.61 & 7.83 & 6.91 &  9.37 & 182 & 0.120 & Minor \\
  DD8, DD9            & $\cdots$ & 1.46 & 7.23529 & 0.08 & 0.35 & 8.10 & 7.70 &  9.73 & 363 & 0.400 & Major \\
  DD10, DD11          & $\cdots$ & 2.84 & 2.34691 & 0.08 & 0.38 & 7.21 & 6.64 &  8.80 & 120 & 0.267 & Major \\
  DD12, DD13          & $\cdots$ & 2.60 & 3.65012 & 0.07 & 0.31 & 7.37 & 7.09 &  9.05 & 175 & 0.526 & Major \\
  DD14, DD15$^\dagger$ & LRD    & 1.40 & 2.41363 & 0.21 & 1.48 & 9.07 & 6.66 & 10.57 & 462 & 0.004 & Satellite \\
  DD16, DD17$^\dagger$ & LRD    & 1.74 & 3.98163 & 0.17 & 0.89 & 6.80 & 7.79 &  9.32 & 143 & 0.103 & Minor \\
  DD18, DD19$^\dagger$ & LRD    & 1.44 & 5.05000 & 0.18 & 0.99 & 9.19 & 8.02 & 10.71 & 667 & 0.068 & Satellite \\
  DD20, DD21$^\dagger$ & LRD    & 1.84 & 6.58456 & 0.15 & 0.62 & 8.64 & 7.96 & 10.22 & 477 & 0.206 & Minor \\
  DD22, DD23$^\dagger$ & LRD    & 1.59 & 6.34000 & 0.09 & 0.39 & 9.14 & 6.00 & 10.63 & 969 & 0.001 & Satellite \\
  DD24, DD25          & $\cdots$ & 1.44 & 2.05996 & 0.19 & 1.32 & 7.37 & 7.01 &  9.02 &  83 & 0.429 & Major \\
  DD26, DD27          & $\cdots$ & 1.44 & 5.65107 & 0.26 & 1.33 & 8.40 & 7.42 &  9.93 & 236 & 0.105 & Minor \\
  DD28, DD29$^\dagger$ & $\cdots$ & 1.65 & 5.14826 & 0.25 & 1.25 & 8.21 & 7.76 &  9.83 & 216 & 0.354 & Major \\
  DD30, DD31$^\dagger$ & $\cdots$ & 2.11 & 3.20763 & 0.18 & 0.96 & 8.72 & 7.87 & 10.27 & 407 & 0.142 & Minor \\
  DD32, DD33          & $\cdots$ & 1.39 & 1.31800 & 0.07 & 0.47 & 7.04 & 6.43 &  8.63 &  88 & 0.241 & Minor \\
  DD34, DD35$^\dagger$ & LRD    & 2.06 & 5.80000 & 0.15 & 0.62 & 8.70 & 8.45 & 10.38 & 578 & 0.558 & Major \\
  DD36, DD37$^\dagger$ & $\cdots$ & 1.78 & 7.87713 & 0.24 & 0.91 & 9.12 & 8.18 & 10.66 & 656 & 0.115 & Minor \\
  DD38, DD39          & $\cdots$ & 1.77 & 6.20331 & 0.23 & 0.98 & 8.63 & 7.31 & 10.14 & 350 & 0.048 & Satellite \\
  DD40, DD41          & $\cdots$ & 1.36 & 3.20855 & 0.12 & 0.81 & 7.79 & 7.13 &  9.37 & 157 & 0.219 & Minor \\
  DD42, DD43          & $\cdots$ & 1.41 & 1.75942 & 0.23 & 1.65 & 7.52 & 7.29 &  9.21 &  92 & 0.588 & Major \\
  DD44, DD45          & $\cdots$ & 1.95 & 1.56558 & 0.10 & 0.62 & 7.07 & 6.58 &  8.68 &  82 & 0.323 & Major \\
  DD46, DD47$^\dagger$ & $\cdots$ & 2.50 & 5.16353 & 0.15 & 0.60 & 8.61 & 6.65 & 10.11 & 430 & 0.011 & Satellite \\
  DD48, DD49$^\dagger$ & $\cdots$ & 1.62 & 3.01792 & 0.08 & 0.50 & 6.00 & 8.73 & 10.22 & 536 & 0.002 & Satellite \\
  DD50, DD51$^\dagger$ & LRD    & 2.70 & 7.04000 & 0.11 & 0.36 & 6.00 & 8.25 &  9.75 & 366 & 0.006 & Satellite \\
  DD52, DD53          & $\cdots$ & 1.82 & 3.14236 & 0.08 & 0.47 & 7.77 & 6.00 &  9.26 & 183 & 0.017 & Satellite \\
  DD54, DD55$^\dagger$ & LRD    & 3.37 & 6.26000 & 0.09 & 0.28 & 8.70 & 8.41 & 10.37 & 840 & 0.511 & Major \\
  DD56, DD57$^\dagger$ & LRD    & 1.59 & 4.96000 & 0.06 & 0.29 & 8.98 & 7.98 & 10.51 & 978 & 0.100 & Satellite \\
  DD58, DD59$^\dagger$ & LRD    & 1.34 & 6.76000 & 0.17 & 0.82 & 8.72 & 7.90 & 10.27 & 445 & 0.153 & Minor \\
  DD60, DD61$^\dagger$ & LRD    & 1.90 & 4.14428 & 0.16 & 0.81 & 8.95 & 7.89 & 10.47 & 564 & 0.087 & Satellite \\
\enddata
\tablecomments{
(1) ``Double Dot" pairs IDs, (2) previously identified class (BLAGN or LRD), 
(3) delensed magnification, (4) redshift ($z_{spec}$, where available, and otherwise $z_{phot}$,
from Table~1), (5) observed separation, (6) delensed physical
separation ($r=r_\mathrm{obs}/\sqrt{\mu}$), (7) and (8) delensed stellar masses of the two members 
($M_\star=M_\star^\mathrm{obs}/\mu$; labeled as ``A" or ``B" in the order of the listed IDs of the members),
(9) total mass, assuming a dark matter ratio of 30, (10) escape velocity,
(11) mass ratio, and (12) merger type based on the mass ratio:
major ($q>0.25$), minor ($0.10<q\le0.25$), and satellite ($q\le0.10$).
$\dagger$At least one member of the pair satisfies the v-shape criterion for an LRD from \citet{kocevski25}.
}
\end{deluxetable}

\clearpage
\bibliography{red_bib}{}

@ARTICLE{barger26,
       author = {{Barger}, A.~J. and {Cowie}, L.~L. and {McKay}, S.~J. and {Bauer}, F.~E.},
        title = "{A Redshift-based Red Selection of Dusty Star-forming Galaxies}",
      journal = {arXiv e-prints},
         year = 2026,
        month = may,
          eid = {arXiv:2605.24264},
        pages = {arXiv:2605.24264},
archivePrefix = {arXiv},
       hideeprint = {2605.24264},
 primaryClass = {astro-ph.GA},
       adsurl = {https://ui.adsabs.harvard.edu/abs/2026arXiv260524264B}
}

@ARTICLE{golubchik26,
       author = {{Golubchik}, Miriam and {Furtak}, Lukas J. and {Allingham}, Joseph F.~V. and {Zitrin}, Adi and {Akins}, Hollis B. and {Kokorev}, Vasily and {Fujimoto}, Seiji and {Abdurro'uf} and {Amor{\'\i}n}, Ricardo O. and {Bauer}, Franz E. and {Bezanson}, Rachel and {Brada{\v{c}}}, Marusa and {Bradley}, Larry D. and {Brammer}, Gabriel B. and {Chisholm}, John and {Coe}, Dan and {Conselice}, Christopher J. and {Dayal}, Pratika and {Dessauges-Zavadsky}, Miroslava and {Diego}, Jose M. and {Faisst}, Andreas L. and {Fei}, Qinyue and {Ferguson}, Henry C. and {Finkelstein}, Steven L. and {Frye}, Brenda L. and {Gonz{\'a}lez-Otero}, Mauro and {Greene}, Jenny E. and {Harikane}, Yuichi and {Hsiao}, Tiger Yu-Yang and {Inayoshi}, Kohei and {Jim{\'e}nez-Teja}, Yolanda and {Knudsen}, Kirsten and {Koekemoer}, Anton M. and {Labb{\'e}}, Ivo and {Lucas}, Ray A. and {Magdis}, Georgios E. and {Matthee}, Jorryt and {Messa}, Matteo and {Naidu}, Rohan P. and {Nakane}, Minami and {Noirot}, Ga{\"e}l and {Pan}, Richard and {Papovich}, Casey and {Richard}, Johan and {Ricotti}, Massimo and {Robbins}, Luke and {Stark}, Daniel P. and {Sun}, Fengwu and {Treu}, Tommaso and {Tripodi}, Roberta and {Vanzella}, Eros and {Willott}, Chris and {Windhorst}, Rogier A.},
        title = "{VENUS: When Red meets Blue -- A multiply imaged Little Red Dot with an apparent blue companion behind the galaxy cluster Abell 383}",
      journal = {arXiv e-prints},
         year = 2025,
        month = dec,
          eid = {arXiv:2512.02117},
        pages = {arXiv:2512.02117},
          hidedoi = {10.48550/arXiv.2512.02117},
archivePrefix = {arXiv},
       hideeprint = {2512.02117},
 primaryClass = {astro-ph.GA},
       adsurl = {https://ui.adsabs.harvard.edu/abs/2025arXiv251202117G}
}

@ARTICLE{loeb94,
       author = {{Loeb}, Abraham and {Rasio}, Frederic A.},
        title = "{Collapse of Primordial Gas Clouds and the Formation of Quasar Black Holes}",
      journal = {\apj},
         year = 1994,
        month = sep,
       volume = {432},
        pages = {52},
          hidedoi = {10.1086/174548},
archivePrefix = {arXiv},
       hideeprint = {astro-ph/9401026},
 primaryClass = {astro-ph},
       adsurl = {https://ui.adsabs.harvard.edu/abs/1994ApJ...432...52L}
}

@ARTICLE{lodato06,
       author = {{Lodato}, Giuseppe and {Natarajan}, Priyamvada},
        title = "{Supermassive black hole formation during the assembly of pre-galactic discs}",
      journal = {\mnras},
         year = 2006,
        month = oct,
       volume = {371},
       number = {4},
        pages = {1813-1823},
          hidedoi = {10.1111/j.1365-2966.2006.10801.x},
archivePrefix = {arXiv},
       hideeprint = {astro-ph/0606159},
 primaryClass = {astro-ph},
       adsurl = {https://ui.adsabs.harvard.edu/abs/2006MNRAS.371.1813L}
}

@ARTICLE{lodato07,
       author = {{Lodato}, Giuseppe and {Natarajan}, Priyamvada},
        title = "{The mass function of high-redshift seed black holes}",
      journal = {\mnras},
         year = 2007,
        month = may,
       volume = {377},
       number = {1},
        pages = {L64-L68},
          hidedoi = {10.1111/j.1745-3933.2007.00304.x},
archivePrefix = {arXiv},
       hideeprint = {astro-ph/0702340},
 primaryClass = {astro-ph},
       adsurl = {https://ui.adsabs.harvard.edu/abs/2007MNRAS.377L..64L}
}

@ARTICLE{begelman06,
       author = {{Begelman}, Mitchell C. and {Volonteri}, Marta and {Rees}, Martin J.},
        title = "{Formation of supermassive black holes by direct collapse in pre-galactic haloes}",
      journal = {\mnras},
         year = 2006,
        month = jul,
       volume = {370},
       number = {1},
        pages = {289-298},
          hidedoi = {10.1111/j.1365-2966.2006.10467.x},
archivePrefix = {arXiv},
       hideeprint = {astro-ph/0602363},
 primaryClass = {astro-ph},
       adsurl = {https://ui.adsabs.harvard.edu/abs/2006MNRAS.370..289B}
}

@ARTICLE{volonteri05,
       author = {{Volonteri}, Marta and {Rees}, Martin J.},
        title = "{Rapid Growth of High-Redshift Black Holes}",
      journal = {\apj},
         year = 2005,
        month = nov,
       volume = {633},
       number = {2},
        pages = {624-629},
          hidedoi = {10.1086/466521},
archivePrefix = {arXiv},
       hideeprint = {astro-ph/0506040},
 primaryClass = {astro-ph},
       adsurl = {https://ui.adsabs.harvard.edu/abs/2005ApJ...633..624V}
}

@ARTICLE{baggen26,
       author = {{Baggen}, Josephine F.~W. and {Scoggins}, Matthew T. and {van Dokkum}, Pieter and {Haiman}, Zolt{\'a}n and {Torralba}, Alberto and {Matthee}, Jorryt},
        title = "{Connecting the Dots: UV-Bright Companions of Little Red Dots as Lyman-Werner Sources Enabling Direct Collapse Black Hole Formation}",
      journal = {arXiv e-prints},
         year = 2026,
        month = feb,
          eid = {arXiv:2602.02702},
        pages = {arXiv:2602.02702},
          hidedoi = {10.48550/arXiv.2602.02702},
archivePrefix = {arXiv},
       hideeprint = {2602.02702},
 primaryClass = {astro-ph.GA},
       adsurl = {https://ui.adsabs.harvard.edu/abs/2026arXiv260202702B}
}

@ARTICLE{pacucci26,
       author = {{Pacucci}, Fabio and {Ferrara}, Andrea and {Kocevski}, Dale D.},
        title = "{The Little Red Dots Are Direct Collapse Black Holes}",
      journal = {arXiv e-prints},
         year = 2026,
        month = jan,
          eid = {arXiv:2601.14368},
        pages = {arXiv:2601.14368},
          hidedoi = {10.48550/arXiv.2601.14368},
archivePrefix = {arXiv},
       hideeprint = {2601.14368},
 primaryClass = {astro-ph.GA},
       adsurl = {https://ui.adsabs.harvard.edu/abs/2026arXiv260114368P}
}

@ARTICLE{pacucci24,
       author = {{Pacucci}, Fabio and {Narayan}, Ramesh},
        title = "{Mildly Super-Eddington Accretion onto Slowly Spinning Black Holes Explains the X-Ray Weakness of the Little Red Dots}",
      journal = {\apj},
         year = 2024,
        month = nov,
       volume = {976},
       number = {1},
          eid = {96},
        pages = {96},
          hidedoi = {10.3847/1538-4357/ad84f7},
archivePrefix = {arXiv},
       hideeprint = {2407.15915},
 primaryClass = {astro-ph.HE},
       adsurl = {https://ui.adsabs.harvard.edu/abs/2024ApJ...976...96P}
}

@ARTICLE{chisolm26,
       author = {{Chisholm}, John and {Berg}, Danielle A. and {Boylan-Kolchin}, Michael and {de Graaff}, Anna and {Furtak}, Lukas J. and {Kokorev}, Vasily and {Matthee}, Jorryt and {Mu{\~n}oz}, Julian B. and {Naidu}, Rohan P. and {Sander}, Andreas A.~C.},
        title = "{Little Red Dots as Globular Clusters in Formation}",
      journal = {arXiv e-prints},
         year = 2026,
        month = feb,
          eid = {arXiv:2602.15935},
        pages = {arXiv:2602.15935},
          hidedoi = {10.48550/arXiv.2602.15935},
archivePrefix = {arXiv},
       hideeprint = {2602.15935},
 primaryClass = {astro-ph.GA},
       adsurl = {https://ui.adsabs.harvard.edu/abs/2026arXiv260215935C}
}

@ARTICLE{maiolino24,
       author = {{Maiolino}, Roberto and {Scholtz}, Jan and {Curtis-Lake}, Emma and {Carniani}, Stefano and {Baker}, William and {de Graaff}, Anna and {Tacchella}, Sandro and {{\"U}bler}, Hannah and {D'Eugenio}, Francesco and {Witstok}, Joris and {Curti}, Mirko and {Arribas}, Santiago and {Bunker}, Andrew J. and {Charlot}, St{\'e}phane and {Chevallard}, Jacopo and {Eisenstein}, Daniel J. and {Egami}, Eiichi and {Ji}, Zhiyuan and {Jones}, Gareth C. and {Lyu}, Jianwei and {Rawle}, Tim and {Robertson}, Brant and {Rujopakarn}, Wiphu and {Perna}, Michele and {Sun}, Fengwu and {Venturi}, Giacomo and {Williams}, Christina C. and {Willott}, Chris},
        title = "{JADES: The diverse population of infant black holes at 4 < z < 11: Merging, tiny, poor, but mighty}",
      journal = {\aap},
         year = 2024,
        month = nov,
       volume = {691},
          eid = {A145},
        pages = {A145},
          hidedoi = {10.1051/0004-6361/202347640},
archivePrefix = {arXiv},
       hideeprint = {2308.01230},
 primaryClass = {astro-ph.GA},
       adsurl = {https://ui.adsabs.harvard.edu/abs/2024A&A...691A.145M}
}

@ARTICLE{mazzolari26,
       author = {{Mazzolari}, G. and {Gilli}, R. and {Maiolino}, R. and {Prandoni}, I. and {Delvecchio}, I. and {Norman}, C. and {Jim{\'e}nez-Andrade}, E.~F. and {Belladitta}, S. and {Vito}, F. and {Momjian}, E. and {Chiaberge}, M. and {Trefoloni}, B. and {Signorini}, M. and {Ji}, X. and {D'Amato}, Q. and {Risaliti}, G. and {Baldi}, R.~D. and {Fabian}, A. and {{\"U}bler}, H. and {D'Eugenio}, F. and {Scholtz}, J. and {Juod{\v{z}}balis}, I. and {Mignoli}, M. and {Brusa}, M. and {Murphy}, E.~J. and {Muxlow}, T.~W.~B.},
        title = "{The radio properties of the JWST-discovered AGN}",
      journal = {\aap},
         year = 2026,
        month = feb,
       volume = {706},
          eid = {A372},
        pages = {A372},
          hidedoi = {10.1051/0004-6361/202453317},
archivePrefix = {arXiv},
       hideeprint = {2412.04224},
 primaryClass = {astro-ph.GA},
       adsurl = {https://ui.adsabs.harvard.edu/abs/2026A&A...706A.372M}
}

@article{degraaff25,
       author = {{de Graaff}, Anna and {Brammer}, Gabriel and {Weibel}, Andrea and {Lewis}, Zach and {Maseda}, Michael V. and {Oesch}, Pascal A. and {Bezanson}, Rachel and {Boogaard}, Leindert A. and {Cleri}, Nikko J. and {Cooper}, Olivia R. and others},
        title = "{RUBIES: A complete census of the bright and red distant universe with JWST/NIRSpec}",
      journal = {Astronomy \& Astrophysics},
         year = 2025,
        month = may,
       volume = {697},
        pages = {A189},
          hidedoi = {10.1051/0004-6361/202452186},
archivePrefix = {arXiv},
       hideeprint = {2409.05948},
 primaryClass = {astro-ph.GA}
}

@article{heintz25,
  title={The JWST-PRIMAL archival survey: A JWST/NIRSpec reference sample for the physical properties and Lyman-$\alpha$ absorption and emission of $\sim$ 600 galaxies at $z= 5.0-13.4$},
  author={Heintz, Kasper E and Brammer, Gabriel B and Watson, Darach and Oesch, Pascal A and Keating, Laura C and Hayes, Matthew J and Abdurro'uf and Arellano-C{\'o}rdova, Karla Z and Carnall, Adam C and Christiansen, Claus R and others},
  journal={Astronomy \& Astrophysics},
  volume={693},
  pages={A60},
  year={2025},
  publisher={EDP Sciences},
  hidedoi={10.1051/0004-6361/202450243}
}

@article{hausen22,
    author = {{Hausen}, Ryan and {Robertson}, Brant E.},
    title = "{FitsMap: A simple, lightweight tool for displaying interactive astronomical image and catalog data}",
    journal = {Astronomy and Computing},
    volume = {39},
    pages = {100586},
    year = {2022},
    month = {April},
    hidedoi = {10.1016/j.ascom.2022.100586},
    archivePrefix = {arXiv},
    hideeprint = {2201.12308},
    primaryClass = {astro-ph.IM}
}

@ARTICLE{harikane23,
       author = {{Harikane}, Yuichi and {Zhang}, Yechi and {Nakajima}, Kimihiko and {Ouchi}, Masami and {Isobe}, Yuki and {Ono}, Yoshiaki and {Hatano}, Shun and {Xu}, Yi and {Umeda}, Hiroya},
        title = "{A JWST/NIRSpec First Census of Broad-line AGNs at z = 4-7: Detection of 10 Faint AGNs with M $_{BH}$ {}10$^{6}$-{}10$^{8}$ M $_{{\ensuremath{\odot}}}$ and Their Host Galaxy Properties}",
      journal = {\apj},
         year = 2023,
        month = dec,
       volume = {959},
       number = {1},
          eid = {39},
        pages = {39},
          hidedoi = {10.3847/1538-4357/ad029e},
archivePrefix = {arXiv},
       hideeprint = {2303.11946},
 primaryClass = {astro-ph.GA},
       adsurl = {https://ui.adsabs.harvard.edu/abs/2023ApJ...959...39H}
}

@INPROCEEDINGS{markwardt09,
  author = {{Markwardt}, C.~B.},
  title = {{Non-linear Least-squares Fitting in IDL with MPFIT}},
  booktitle = {Astronomical Data Analysis Software and Systems XVIII},
  year = {2009},
  editor = {{Bohlender}, D.~A. and {Durand}, D. and {Dowler}, P.},
  volume = {411},
  series = {Astronomical Society of the Pacific Conference Series},
  pages = {251},
  month = sep,
  adsurl = {http://ads.ari.uni-heidelberg.de/abs/2009ASPC..411..251M},
  archiveprefix = {arXiv},
  hideeprint = {0902.2850},
  hideprimaryclass = {astro-ph.IM}
}

@ARTICLE{ananna24,
       author = {{Ananna}, Tonima Tasnim and {Bogd{\'a}n}, {\'A}kos and {Kov{\'a}cs}, Orsolya E. and {Natarajan}, Priyamvada and {Hickox}, Ryan C.},
        title = "{X-Ray View of Little Red Dots: Do They Host Supermassive Black Holes?}",
      journal = {\apjl},
         year = 2024,
        month = jul,
       volume = {969},
       number = {1},
          eid = {L18},
        pages = {L18},
          hidedoi = {10.3847/2041-8213/ad5669},
archivePrefix = {arXiv},
       hideeprint = {2404.19010},
 primaryClass = {astro-ph.GA},
       adsurl = {https://ui.adsabs.harvard.edu/abs/2024ApJ...969L..18A}
}

@ARTICLE{larson11,
       author = {{Larson}, D. and {Dunkley}, J. and {Hinshaw}, G. and {Komatsu}, E. and {Nolta}, M.~R. and {Bennett}, C.~L. and {Gold}, B. and {Halpern}, M. and {Hill}, R.~S. and {Jarosik}, N. and {Kogut}, A. and {Limon}, M. and {Meyer}, S.~S. and {Odegard}, N. and {Page}, L. and {Smith}, K.~M. and {Spergel}, D.~N. and {Tucker}, G.~S. and {Weiland}, J.~L. and {Wollack}, E. and {Wright}, E.~L.},
        title = "{Seven-year Wilkinson Microwave Anisotropy Probe (WMAP) Observations: Power Spectra and WMAP-derived Parameters}",
      journal = {\apjs},
         year = 2011,
        month = feb,
       volume = {192},
       number = {2},
          eid = {16},
        pages = {16},
          hidedoi = {10.1088/0067-0049/192/2/16},
archivePrefix = {arXiv},
       hideeprint = {1001.4635},
 primaryClass = {astro-ph.CO},
       adsurl = {https://ui.adsabs.harvard.edu/abs/2011ApJS..192...16L}
}

@article{bezanson22,
 adsurl = {https://ui.adsabs.harvard.edu/abs/2024ApJ...974...92B},
 archiveprefix = {arXiv},
 author = {{Bezanson}, Rachel and {Labbe}, Ivo and {Whitaker}, Katherine E. and {Leja}, Joel and {Price}, Sedona H. and {Franx}, Marijn and {Brammer}, Gabriel and {Marchesini}, Danilo and {Zitrin}, Adi and {Wang}, Bingjie and {Weaver}, John R. and {Furtak}, Lukas J. and {Atek}, Hakim and {Coe}, Dan and {Cutler}, Sam E. and {Dayal}, Pratika and {van Dokkum}, Pieter and {Feldmann}, Robert and {F{\"o}rster Schreiber}, Natascha M. and {Fujimoto}, Seiji and {Geha}, Marla and {Glazebrook}, Karl and {de Graaff}, Anna and {Greene}, Jenny E. and {Juneau}, St{\'e}phanie and {Kassin}, Susan and {Kriek}, Mariska and {Khullar}, Gourav and {Maseda}, Michael and {Mowla}, Lamiya A. and {Muzzin}, Adam and {Nanayakkara}, Themiya and {Nelson}, Erica J. and {Oesch}, Pascal A. and {Pacifici}, Camilla and {Pan}, Richard and {Papovich}, Casey and {Setton}, David J. and {Shapley}, Alice E. and {Smit}, Renske and {Stefanon}, Mauro and {Taylor}, Edward N. and {Williams}, Christina C.},
 eid = {92},
 hidedoi = {10.3847/1538-4357/ad66cf},
 hideeprint = {2212.04026},
 journal = {\apj},
 month = {October},
 number = {1},
 pages = {92},
 primaryclass = {astro-ph.GA},
 title = {{The JWST UNCOVER Treasury Survey: Ultradeep NIRSpec and NIRCam Observations before the Epoch of Reionization}},
 volume = {974},
 year = {2024}
}

@article{furtak23,
 adsurl = {https://ui.adsabs.harvard.edu/abs/2023ApJ...952..142F},
 archiveprefix = {arXiv},
 author = {{Furtak}, Lukas J. and {Zitrin}, Adi and {Plat}, Ad{\`e}le and {Fujimoto}, Seiji and {Wang}, Bingjie and {Nelson}, Erica J. and {Labb{\'e}}, Ivo and {Bezanson}, Rachel and {Brammer}, Gabriel B. and {van Dokkum}, Pieter and {Endsley}, Ryan and {Glazebrook}, Karl and {Greene}, Jenny E. and {Leja}, Joel and {Price}, Sedona H. and {Smit}, Renske and {Stark}, Daniel P. and {Weaver}, John R. and {Whitaker}, Katherine E. and {Atek}, Hakim and {Chevallard}, Jacopo and {Curtis-Lake}, Emma and {Dayal}, Pratika and {Feltre}, Anna and {Franx}, Marijn and {Fudamoto}, Yoshinobu and {Marchesini}, Danilo and {Mowla}, Lamiya A. and {Pan}, Richard and {Suess}, Katherine A. and {Vidal-Garc{\'\i}a}, Alba and {Williams}, Christina C.},
 eid = {142},
 hidedoi = {10.3847/1538-4357/acdc9d},
 hideeprint = {2212.10531},
 journal = {\apj},
 month = {August},
 number = {2},
 pages = {142},
 primaryclass = {astro-ph.GA},
 title = {{JWST UNCOVER: Extremely Red and Compact Object at z $_{phot}$ \UTF{2243} 7.6 Triply Imaged by A2744}},
 volume = {952},
 year = {2023}
}

@article{furtak23b,
 adsurl = {https://ui.adsabs.harvard.edu/abs/2023MNRAS.523.4568F},
 archiveprefix = {arXiv},
 author = {{Furtak}, Lukas J. and {Zitrin}, Adi and {Weaver}, John R. and {Atek}, Hakim and {Bezanson}, Rachel and {Labb{\'e}}, Ivo and {Whitaker}, Katherine E. and {Leja}, Joel and {Price}, Sedona H. and {Brammer}, Gabriel B. and {Wang}, Bingjie and {Marchesini}, Danilo and {Pan}, Richard and {Dayal}, Pratika and {van Dokkum}, Pieter and {Feldmann}, Robert and {Fujimoto}, Seiji and {Franx}, Marijn and {Khullar}, Gourav and {Nelson}, Erica J. and {Mowla}, Lamiya A.},
 hidedoi = {10.1093/mnras/stad1627},
 hideeprint = {2212.04381},
 journal = {\mnras},
 month = {August},
 number = {3},
 pages = {4568-4582},
 primaryclass = {astro-ph.GA},
 title = {{UNCOVERing the extended strong lensing structures of Abell 2744 with the deepest JWST imaging}},
 volume = {523},
 year = {2023}
}

@article{greene24,
 adsurl = {https://ui.adsabs.harvard.edu/abs/2024ApJ...964...39G},
 archiveprefix = {arXiv},
 author = {{Greene}, Jenny E. and {Labbe}, Ivo and {Goulding}, Andy D. and {Furtak}, Lukas J. and {Chemerynska}, Iryna and {Kokorev}, Vasily and {Dayal}, Pratika and {Volonteri}, Marta and {Williams}, Christina C. and {Wang}, Bingjie and {Setton}, David J. and {Burgasser}, Adam J. and {Bezanson}, Rachel and {Atek}, Hakim and {Brammer}, Gabriel and {Cutler}, Sam E. and {Feldmann}, Robert and {Fujimoto}, Seiji and {Glazebrook}, Karl and {de Graaff}, Anna and {Khullar}, Gourav and {Leja}, Joel and {Marchesini}, Danilo and {Maseda}, Michael V. and {Matthee}, Jorryt and {Miller}, Tim B. and {Naidu}, Rohan P. and {Nanayakkara}, Themiya and {Oesch}, Pascal A. and {Pan}, Richard and {Papovich}, Casey and {Price}, Sedona H. and {van Dokkum}, Pieter and {Weaver}, John R. and {Whitaker}, Katherine E. and {Zitrin}, Adi},
 eid = {39},
 hidedoi = {10.3847/1538-4357/ad1e5f},
 hideeprint = {2309.05714},
 journal = {\apj},
 month = {March},
 number = {1},
 pages = {39},
 primaryclass = {astro-ph.GA},
 title = {{UNCOVER Spectroscopy Confirms the Surprising Ubiquity of Active Galactic Nuclei in Red Sources at z > 5}},
 volume = {964},
 year = {2024}
}

@ARTICLE{kocevski25,
       author = {{Kocevski}, Dale D. and {Finkelstein}, Steven L. and {Barro}, Guillermo and {Taylor}, Anthony J. and {Calabr{\`o}}, Antonello and {Laloux}, Brivael and {Buchner}, Johannes and {Trump}, Jonathan R. and {Leung}, Gene C.~K. and {Yang}, Guang and {Dickinson}, Mark and {P{\'e}rez-Gonz{\'a}lez}, Pablo G. and {Pacucci}, Fabio and {Inayoshi}, Kohei and {Somerville}, Rachel S. and {McGrath}, Elizabeth J. and {Akins}, Hollis B. and {Bagley}, Micaela B. and {Bowler}, Rebecca A.~A. and {Bisigello}, Laura and {Carnall}, Adam and {Casey}, Caitlin M. and {Cheng}, Yingjie and {Cleri}, Nikko J. and {Costantin}, Luca and {Cullen}, Fergus and {Davis}, Kelcey and {Donnan}, Callum T. and {Dunlop}, James S. and {Ellis}, Richard S. and {Ferguson}, Henry C. and {Fujimoto}, Seiji and {Fontana}, Adriano and {Giavalisco}, Mauro and {Grazian}, Andrea and {Grogin}, Norman A. and {Hathi}, Nimish P. and {Hirschmann}, Michaela and {Huertas-Company}, Marc and {Holwerda}, Benne W. and {Illingworth}, Garth and {Juneau}, St{\'e}phanie and {Kartaltepe}, Jeyhan S. and {Koekemoer}, Anton M. and {Li}, Wenxiu and {Lucas}, Ray A. and {Magee}, Dan and {Mason}, Charlotte and {McLeod}, Derek J. and {McLure}, Ross J. and {Napolitano}, Lorenzo and {Papovich}, Casey and {Pirzkal}, Nor and {Rodighiero}, Giulia and {Santini}, Paola and {Wilkins}, Stephen M. and {Yung}, L.~Y. Aaron},
        title = "{The Rise of Faint, Red Active Galactic Nuclei at z > 4: A Sample of Little Red Dots in the JWST Extragalactic Legacy Fields}",
      journal = {\apj},
         year = 2025,
        month = jun,
       volume = {986},
       number = {2},
          eid = {126},
        pages = {126},
          hidedoi = {10.3847/1538-4357/adbc7d},
archivePrefix = {arXiv},
       hideeprint = {2404.03576},
 primaryClass = {astro-ph.GA},
       adsurl = {https://ui.adsabs.harvard.edu/abs/2025ApJ...986..126K}
}

@article{kroupa01,
 adsurl = {https://ui.adsabs.harvard.edu/abs/2001MNRAS.322..231K},
 archiveprefix = {arXiv},
 author = {{Kroupa}, Pavel},
 hidedoi = {10.1046/j.1365-8711.2001.04022.x},
 hideeprint = {astro-ph/0009005},
 journal = {\mnras},
 month = {April},
 number = {2},
 pages = {231-246},
 primaryclass = {astro-ph},
 title = {{On the variation of the initial mass function}},
 volume = {322},
 year = {2001}
}

@ARTICLE{labbe25,
       author = {{Labbe}, Ivo and {Greene}, Jenny E. and {Bezanson}, Rachel and {Fujimoto}, Seiji and {Furtak}, Lukas J. and {Goulding}, Andy D. and {Matthee}, Jorryt and {Naidu}, Rohan P. and {Oesch}, Pascal A. and {Atek}, Hakim and {Brammer}, Gabriel and {Chemerynska}, Iryna and {Coe}, Dan and {Cutler}, Sam E. and {Dayal}, Pratika and {Feldmann}, Robert and {Franx}, Marijn and {Glazebrook}, Karl and {Leja}, Joel and {Maseda}, Michael and {Marchesini}, Danilo and {Nanayakkara}, Themiya and {Nelson}, Erica J. and {Pan}, Richard and {Papovich}, Casey and {Price}, Sedona H. and {Suess}, Katherine A. and {Wang}, Bingjie and {Weaver}, John R. and {Whitaker}, Katherine E. and {Williams}, Christina C. and {Zitrin}, Adi},
        title = "{UNCOVER: Candidate Red Active Galactic Nuclei at 3 < z < 7 with JWST and ALMA}",
      journal = {\apj},
         year = 2025,
        month = jan,
       volume = {978},
       number = {1},
          eid = {92},
        pages = {92},
         hidedoi = {10.3847/1538-4357/ad3551},
archivePrefix = {arXiv},
       hideeprint = {2306.07320},
 primaryClass = {astro-ph.GA},
       adsurl = {https://ui.adsabs.harvard.edu/abs/2025ApJ...978...92L}
}

@article{matthee24,
 adsurl = {https://ui.adsabs.harvard.edu/abs/2024ApJ...963..129M},
 archiveprefix = {arXiv},
 author = {{Matthee}, Jorryt and {Naidu}, Rohan P. and {Brammer}, Gabriel and {Chisholm}, John and {Eilers}, Anna-Christina and {Goulding}, Andy and {Greene}, Jenny and {Kashino}, Daichi and {Labbe}, Ivo and {Lilly}, Simon J. and {Mackenzie}, Ruari and {Oesch}, Pascal A. and {Weibel}, Andrea and {Wuyts}, Stijn and {Xiao}, Mengyuan and {Bordoloi}, Rongmon and {Bouwens}, Rychard and {van Dokkum}, Pieter and {Illingworth}, Garth and {Kramarenko}, Ivan and {Maseda}, Michael V. and {Mason}, Charlotte and {Meyer}, Romain A. and {Nelson}, Erica J. and {Reddy}, Naveen A. and {Shivaei}, Irene and {Simcoe}, Robert A. and {Yue}, Minghao},
 eid = {129},
 hidedoi = {10.3847/1538-4357/ad2345},
 hideeprint = {2306.05448},
 journal = {\apj},
 month = {March},
 number = {2},
 pages = {129},
 primaryclass = {astro-ph.GA},
 title = {{Little Red Dots: An Abundant Population of Faint Active Galactic Nuclei at z {\ensuremath{\sim}} 5 Revealed by the EIGER and FRESCO JWST Surveys}},
 volume = {963},
 year = {2024}
}

@ARTICLE{taylor25,
       author = {{Taylor}, Anthony J. and {Finkelstein}, Steven L. and {Kocevski}, Dale D. and {Jeon}, Junehyoung and {Bromm}, Volker and {Amor{\'\i}n}, Ricardo O. and {Arrabal Haro}, Pablo and {Backhaus}, Bren E. and {Bagley}, Micaela B. and {Banados}, Eduardo and {Bhatawdekar}, Rachana and {Brooks}, Madisyn and {Calabr{\`o}}, Antonello and {Ch{\'a}vez Ortiz}, {\'O}scar A. and {Cheng}, Yingjie and {Cleri}, Nikko J. and {Cole}, Justin W. and {Davis}, Kelcey and {Dickinson}, Mark and {Donnan}, Callum and {Dunlop}, James S. and {Ellis}, Richard S. and {Fern{\'a}ndez}, Vital and {Fontana}, Adriano and {Fujimoto}, Seiji and {Giavalisco}, Mauro and {Grazian}, Andrea and {Guo}, Jingsong and {Hathi}, Nimish P. and {Holwerda}, Benne W. and {Hirschmann}, Michaela and {Inayoshi}, Kohei and {Kartaltepe}, Jeyhan S. and {Khusanova}, Yana and {Koekemoer}, Anton M. and {Kokorev}, Vasily and {Larson}, Rebecca L. and {Leung}, Gene C.~K. and {Lucas}, Ray A. and {McLeod}, Derek J. and {Napolitano}, Lorenzo and {Onoue}, Masafusa and {Pacucci}, Fabio and {Papovich}, Casey and {P{\'e}rez-Gonz{\'a}lez}, Pablo G. and {Pirzkal}, Nor and {Somerville}, Rachel S. and {Trump}, Jonathan R. and {Wilkins}, Stephen M. and {Yung}, L.~Y. Aaron and {Zhang}, Haowen},
        title = "{Broad-line AGNs at 3.5 < z < 6: The Black Hole Mass Function and a Connection with Little Red Dots}",
      journal = {\apj},
         year = 2025,
        month = jun,
       volume = {986},
       number = {2},
          eid = {165},
        pages = {165},
          hidedoi = {10.3847/1538-4357/add15b},
archivePrefix = {arXiv},
       hideeprint = {2409.06772},
 primaryClass = {astro-ph.GA},
       adsurl = {https://ui.adsabs.harvard.edu/abs/2025ApJ...986..165T}
}

@article{treu22,
 adsurl = {https://ui.adsabs.harvard.edu/abs/2022ApJ...935..110T},
 archiveprefix = {arXiv},
 author = {{Treu}, T. and {Roberts-Borsani}, G. and {Bradac}, M. and {Brammer}, G. and {Fontana}, A. and {Henry}, A. and {Mason}, C. and {Morishita}, T. and {Pentericci}, L. and {Wang}, X. and {Acebron}, A. and {Bagley}, M. and {Bergamini}, P. and {Belfiori}, D. and {Bonchi}, A. and {Boyett}, K. and {Boutsia}, K. and {Calabr{\'o}}, A. and {Caminha}, G.~B. and {Castellano}, M. and {Dressler}, A. and {Glazebrook}, K. and {Grillo}, C. and {Jacobs}, C. and {Jones}, T. and {Kelly}, P.~L. and {Leethochawalit}, N. and {Malkan}, M.~A. and {Marchesini}, D. and {Mascia}, S. and {Mercurio}, A. and {Merlin}, E. and {Nanayakkara}, T. and {Nonino}, M. and {Paris}, D. and {Poggianti}, B. and {Rosati}, P. and {Santini}, P. and {Scarlata}, C. and {Shipley}, H.~V. and {Strait}, V. and {Trenti}, M. and {Tubthong}, C. and {Vanzella}, E. and {Vulcani}, B. and {Yang}, L.},
 eid = {110},
 hidedoi = {10.3847/1538-4357/ac8158},
 hideeprint = {2206.07978},
 journal = {\apj},
 month = {August},
 number = {2},
 pages = {110},
 primaryclass = {astro-ph.GA},
 title = {{The GLASS-JWST Early Release Science Program. I. Survey Design and Release Plans}},
 volume = {935},
 year = {2022}
}

@article{weaver24,
 adsurl = {https://ui.adsabs.harvard.edu/abs/2024ApJS..270....7W},
 archiveprefix = {arXiv},
 author = {{Weaver}, John R. and {Cutler}, Sam E. and {Pan}, Richard and {Whitaker}, Katherine E. and {Labb{\'e}}, Ivo and {Price}, Sedona H. and {Bezanson}, Rachel and {Brammer}, Gabriel and {Marchesini}, Danilo and {Leja}, Joel and {Wang}, Bingjie and {Furtak}, Lukas J. and {Zitrin}, Adi and {Atek}, Hakim and {Chemerynska}, Iryna and {Coe}, Dan and {Dayal}, Pratika and {van Dokkum}, Pieter and {Feldmann}, Robert and {F{\"o}rster Schreiber}, Natascha M. and {Franx}, Marijn and {Fujimoto}, Seiji and {Fudamoto}, Yoshinobu and {Glazebrook}, Karl and {de Graaff}, Anna and {Greene}, Jenny E. and {Juneau}, St{\'e}phanie and {Kassin}, Susan and {Kriek}, Mariska and {Khullar}, Gourav and {Maseda}, Michael V. and {Mowla}, Lamiya A. and {Muzzin}, Adam and {Nanayakkara}, Themiya and {Nelson}, Erica J. and {Oesch}, Pascal A. and {Pacifici}, Camilla and {Papovich}, Casey and {Setton}, David J. and {Shapley}, Alice E. and {Shipley}, Heath V. and {Smit}, Renske and {Stefanon}, Mauro and {Taylor}, Edward N. and {Weibel}, Andrea and {Williams}, Christina C.},
 eid = {7},
 hidedoi = {10.3847/1538-4365/ad07e0},
 hideeprint = {2301.02671},
 journal = {\apjs},
 month = {January},
 number = {1},
 pages = {7},
 primaryclass = {astro-ph.GA},
 title = {{The UNCOVER Survey: A First-look HST + JWST Catalog of 60,000 Galaxies near A2744 and beyond}},
 volume = {270},
 year = {2024}
}
\bibliographystyle{aasjournalv7}

\begin{acknowledgements}
{
We thank Anthony Taylor for helpful discussions.
We gratefully acknowledge support for this research from the
University of Wisconsin-Madison, Office of the Vice Chancellor for Research 
with funding from the Wisconsin Alumni Research Foundation (A.~J.~B.) and
NASA grant 80NSSC22K0483 (L.~L.~C.).
}
\end{acknowledgements}

\facilities{JWST}

\end{document}